\documentclass{hep99}
\usepackage{times}

\usepackage{epsfig}
\usepackage{amsmath}
\usepackage{hhline}
\usepackage{amssymb}
\usepackage{times}
\usepackage{cite}

\begin{document}

%
%
\newcommand{\s}{\mbox{$s$}}
\newcommand{\ttra}{\mbox{$t$}}
\newcommand{\modt}{\mbox{$|t|$}}
\newcommand{\eminpz}{\mbox{$E-p_z$}}
\newcommand{\eminpzs}{\mbox{$\Sigma(E-p_z)$}}
\newcommand{\rap}{\ensuremath{\eta^*} }
\newcommand{\W}{\mbox{$W$}}
\newcommand{\w}{\mbox{$W$}}
\newcommand{\Q}{\mbox{$Q$}}
\newcommand{\q}{\mbox{$Q$}}
\newcommand{\xB}{\mbox{$x$}}  
\newcommand{\xF}{\mbox{$x_F$}}  
\newcommand{\xg}{\mbox{$x_g$}}  
\newcommand{\xbj}{$x$}
\newcommand{\xpom}{x_{I\!\!P}}
\newcommand{\zpom}{z_{I\!\!P}}
\newcommand{\y}{\mbox{$y~$}}
\newcommand{\Qsq}{\mbox{$Q^2$}}
\newcommand{\qzsq}{\mbox{$Q_o^2$}}
\newcommand{\qsq}{\mbox{$Q^2$}}
\newcommand{\kjet}{\mbox{$k_{T\rm{jet}}$}}
\newcommand{\xjet}{\mbox{$x_{\rm{jet}}$}}
\newcommand{\Ejet}{\mbox{$E_{\rm{jet}}$}}
\newcommand{\thjet}{\mbox{$\theta_{\rm{jet}}$}}
\newcommand{\pjet}{\mbox{$p_{T\rm{jet}}$}}
\newcommand{\et}{\mbox{$E_T$}}
\newcommand{\kt}{\mbox{$k_T$}}
\newcommand{\ptrans}{\mbox{$p_T~$}}
\newcommand{\pth}{\mbox{$p_T^h~$}}
\newcommand{\pte}{\mbox{$p_T^e~$}}
\newcommand{\ptsq}{\mbox{$p_T^{\star 2}~$}}
\newcommand{\as}{\mbox{$\alpha_s$}}
\newcommand{\ycut}{\mbox{$y_{\rm cut}~$}}
\newcommand{\gx}{\mbox{$g(x_g,Q^2)$~}}
\newcommand{\xpart}{\mbox{$x_{\rm part~}$}}
\newcommand{\mrsdm}{\mbox{${\rm MRSD}^-~$}}
\newcommand{\mrsdmp}{\mbox{${\rm MRSD}^{-'}~$}}
\newcommand{\mrsdn}{\mbox{${\rm MRSD}^0~$}}
\newcommand{\lambdams}{\mbox{$\Lambda_{\rm \bar{MS}}~$}}
%
%
\newcommand{\gp}{\ensuremath{\gamma}p }
\newcommand{\gammasp}{\ensuremath{\gamma}*p }
\newcommand{\gammap}{\ensuremath{\gamma}p }
\newcommand{\gsp}{\ensuremath{\gamma^*}p }
\newcommand{\dsiget}{\ensuremath{{\rm d}\sigma_{ep}/{\rm d}E_t^*} }
\newcommand{\dsigrap}{\ensuremath{{\rm d}\sigma_{ep}/{\rm d}\eta^*} }
\newcommand{\epem}{\mbox{$e^+e^-$}}
\newcommand{\ep}{\mbox{$ep~$}}
\newcommand{\epl}{\mbox{$e^{+}$}}
\newcommand{\emi}{\mbox{$e^{-}$}}
\newcommand{\epm}{\mbox{$e^{\pm}$}}
\newcommand{\xsec}{cross section}
\newcommand{\xsecs}{cross sections}
\newcommand{\inter}{interaction}
\newcommand{\inters}{interactions}
\newcommand{\eplp}{\mbox{$e^+p$}}
\newcommand{\emip}{\mbox{$e^-p$}}
\newcommand{\gamm}{$\gamma$}
\newcommand{\zzo}{$Z^{o}$}
%
%
\newcommand{\phib}{\mbox{$\varphi$}}
\newcommand{\rh}{\mbox{$\rho$}}
\newcommand{\rhz}{\mbox{$\rh^0$}}
\newcommand{\ph}{\mbox{$\phi$}}
\newcommand{\om}{\mbox{$\omega$}}
\newcommand{\jpsi}{\mbox{$J/\psi$}}
\newcommand{\pipi}{\mbox{$\pi^+\pi^-$}}
\newcommand{\pip}{\mbox{$\pi^+$}}
\newcommand{\pim}{\mbox{$\pi^-$}}
\newcommand{\kk}{\mbox{K^+K^-$}}
\newcommand{\bsl}{\mbox{$b$}}
\newcommand{\alp}{\mbox{$\alpha^\prime$}}
\newcommand{\alpom}{\mbox{$\alpha_{\PO}$}}
\newcommand{\alregg}{\mbox{$\alpha_{\regg}$}}
\newcommand{\alpomp}{\mbox{$\alpha_{\PO}^\prime$}}
\newcommand{\rzzzz}{\mbox{$r_{00}^{04}$}}
\newcommand{\rzqzz}{\mbox{$r_{00}^{04}$}}
\newcommand{\rzquz}{\mbox{$r_{10}^{04}$}}
\newcommand{\rzqumu}{\mbox{$r_{1-1}^{04}$}}
\newcommand{\ruuu}{\mbox{$r_{11}^{1}$}}
\newcommand{\ruzz}{\mbox{$r_{00}^{1}$}}
\newcommand{\ruuz}{\mbox{$r_{10}^{1}$}}
\newcommand{\ruumu}{\mbox{$r_{1-1}^{1}$}}
\newcommand{\rduz}{\mbox{$r_{10}^{2}$}}
\newcommand{\rdumu}{\mbox{$r_{1-1}^{2}$}}
\newcommand{\rcuu}{\mbox{$r_{11}^{5}$}}
\newcommand{\rczz}{\mbox{$r_{00}^{5}$}}
\newcommand{\rcuz}{\mbox{$r_{10}^{5}$}}
\newcommand{\rcumu}{\mbox{$r_{1-1}^{5}$}}
\newcommand{\rsuz}{\mbox{$r_{10}^{6}$}}
\newcommand{\rsumu}{\mbox{$r_{1-1}^{6}$}}
\newcommand{\rzqik}{\mbox{$r_{ik}^{04}$}}
\newcommand{\rhzik}{\mbox{$\rh_{ik}^{0}$}}
\newcommand{\rhqik}{\mbox{$\rh_{ik}^{4}$}}
\newcommand{\rhaik}{\mbox{$\rh_{ik}^{\alpha}$}}
\newcommand{\rhzzz}{\mbox{$\rh_{00}^{0}$}}
\newcommand{\rhqzz}{\mbox{$\rh_{00}^{4}$}}
\newcommand{\raik}{\mbox{$r_{ik}^{\alpha}$}}
\newcommand{\razz}{\mbox{$r_{00}^{\alpha}$}}
\newcommand{\rauz}{\mbox{$r_{10}^{\alpha}$}}
\newcommand{\raumu}{\mbox{$r_{1-1}^{\alpha}$}}

\newcommand{\R}{\mbox{$R$}}
\newcommand{\rzero}{\mbox{$r_{00}^{04}$}}
\newcommand{\rone}{\mbox{$r_{1-1}^{1}$}}
\newcommand{\costh}{\mbox{$\cos\theta$}}
\newcommand{\cosp}{\mbox{$\cos\psi$}}
\newcommand{\costop}{\mbox{$\cos(2\psi)$}}
\newcommand{\cosd}{\mbox{$\cos\delta$}}
\newcommand{\cossqp}{\mbox{$\cos^2\psi$}}
\newcommand{\cossqt}{\mbox{$\cos^2\theta^*$}}
\newcommand{\sint}{\mbox{$\sin\theta^*$}}
\newcommand{\sintot}{\mbox{$\sin(2\theta^*)$}}
\newcommand{\sinsqt}{\mbox{$\sin^2\theta^*$}}
\newcommand{\costhst}{\mbox{$\cos\theta^*$}}
\newcommand{\vep}{\mbox{$V p$}}
\newcommand{\mpipi}{\mbox{$m_{\pi^+\pi^-}$}}
\newcommand{\mkk}{\mbox{$m_{KK}$}}
\newcommand{\mkaka}{\mbox{$m_{K^+K^-}$}}
\newcommand{\mpp}{\mbox{$m_{\pi\pi}$}}       
\newcommand{\mppsq}{\mbox{$m_{\pi\pi}^2$}}   
\newcommand{\mpi}{\mbox{$m_{\pi}$}}          
\newcommand{\mrho}{\mbox{$m_{\rho}$}}        
\newcommand{\mrhosq}{\mbox{$m_{\rho}^2$}}    
\newcommand{\Gmpp}{\mbox{$\Gamma (\mpp)$}}   
\newcommand{\Gmppsq}{\mbox{$\Gamma^2(\mpp)$}}
\newcommand{\Grho}{\mbox{$\Gamma_{\rho}$}}   
\newcommand{\grho}{\mbox{$\Gamma_{\rho}$}}   
\newcommand{\Grhosq}{\mbox{$\Gamma_{\rho}^2$}}   
%
%
\newcommand{\cm}{\mbox{\rm cm}}
\newcommand{\GeV}{\mbox{\rm GeV}}
\newcommand{\gev}{\mbox{\rm GeV}}
\newcommand{\GeVx}{\rm GeV}
\newcommand{\gevx}{\rm GeV}
\newcommand{\GeVc}{\rm GeV/c}
\newcommand{\gevc}{\rm GeV/c}
\newcommand{\MeVc}{\rm MeV/c}
\newcommand{\mevc}{\rm MeV/c}
\newcommand{\MeV}{\mbox{\rm MeV}}
\newcommand{\mev}{\mbox{\rm MeV}}
\newcommand{\MeVx}{\mbox{\rm MeV}}
\newcommand{\mevx}{\mbox{\rm MeV}}
\newcommand{\GeVsq}{\mbox{${\rm GeV}^2$}}
\newcommand{\gevsq}{\mbox{${\rm GeV}^2$}}
\newcommand{\gevsqc}{\mbox{${\rm GeV^2/c^4}$}}
\newcommand{\gevcsq}{\mbox{${\rm GeV/c^2}$}}
\newcommand{\mevcsq}{\mbox{${\rm MeV/c^2}$}}
\newcommand{\GeVsqm}{\mbox{${\rm GeV}^{-2}$}}
\newcommand{\gevsqm}{\mbox{${\rm GeV}^{-2}$}}
\newcommand{\nb}{\mbox{${\rm nb}$}}
\newcommand{\nbinv}{\mbox{${\rm nb^{-1}}$}}
\newcommand{\pbinv}{\mbox{${\rm pb^{-1}}$}}
\newcommand{\mm}{\mbox{$\cdot 10^{-2}$}}
\newcommand{\mmm}{\mbox{$\cdot 10^{-3}$}}
\newcommand{\mmmm}{\mbox{$\cdot 10^{-4}$}}
\newcommand{\degr}{\mbox{$^{\circ}$}}
%
%
\newcommand{\F}{$ F_{2}(x,Q^2)\,$}  
\newcommand{\Fc}{$ F_{2}\,$}    
\newcommand{\XP}{x_{{I\!\!P}/{p}}}       
\newcommand{\TOSS}{x_{{i}/{\PO}}}        
\newcommand{\un}[1]{\mbox{\rm #1}} 
\newcommand{\LO}{Leading Order}
\newcommand{\NLO}{Next to Leading Order}
\newcommand{\ft}{$ F_{2}\,$}
%
%
\newcommand{\mc}{\multicolumn}
\newcommand{\bce}{\begin{center}}
\newcommand{\ece}{\end{center}}
\newcommand{\beq}{\begin{equation}}
\newcommand{\eeq}{\end{equation}}
\newcommand{\bea}{\begin{eqnarray}}
\newcommand{\eea}{\end{eqnarray}}
%
%
\def\lsim{\mathrel{\rlap{\lower4pt\hbox{\hskip1pt$\sim$}}
    \raise1pt\hbox{$<$}}}         
\def\gsim{\mathrel{\rlap{\lower4pt\hbox{\hskip1pt$\sim$}}
    \raise1pt\hbox{$>$}}}         
%
%
\newcommand{\pom}{{I\!\!P}}
\newcommand{\regg}{{I\!\!R}}
\newcommand{\PO}{I\!\!P}
\newcommand{\slowpi}{\pi_{\mathit{slow}}}
\newcommand{\fiidiii}{F_2^{D(3)}}
\newcommand{\fiidiiiarg}{F_2^{D(3)}\,(\beta,\,Q^2,\,x)}
\newcommand{\fiidiiifull}{F_2^{D(3)}\,(x_{I\!\!P},\,\beta,\,Q^2)}
\newcommand{\n}{1.19\pm 0.06 (stat.) \pm0.07 (syst.)}
\newcommand{\nz}{1.30\pm 0.08 (stat.)^{+0.08}_{-0.14} (syst.)}
\newcommand{\fiidiiiifull}{$F_2^{D(4)}\,(\beta,\,Q^2,\,x,\,t)$}
\newcommand{\fiipom}{\tilde F_2^D}
\newcommand{\fiipomfull}{\tilde F_2^D\,(\beta,\,Q^2)}
\newcommand{\ALPHA}{1.10\pm0.03 (stat.) \pm0.04 (syst.)}
\newcommand{\ALPHAZ}{1.15\pm0.04 (stat.)^{+0.04}_{-0.07} (syst.)}
\newcommand{\fiipomarg}{\fiipom\,(\beta,\,Q^2)}
\newcommand{\pomflux}{f_{\pom / p}}
\newcommand{\nxpom}{1.19\pm 0.06 (stat.) \pm0.07 (syst.)}
\newcommand {\gapprox}
   {\raisebox{-0.7ex}{$\stackrel {\textstyle>}{\sim}$}}
\newcommand {\lapprox}
   {\raisebox{-0.7ex}{$\stackrel {\textstyle<}{\sim}$}}
\newcommand{\pomfluxarg}{f_{\pom / p}\,(x_\pom)}
\newcommand{\dsf}{\mbox{$F_2^{D(3)}$}}
\newcommand{\dsfva}{\mbox{$F_2^{D(3)}(\beta,Q^2,x_{I\!\!P})$}}
\newcommand{\dsfvb}{\mbox{$F_2^{D(3)}(\beta,Q^2,x)$}}
\newcommand{\dsfpom}{$F_2^{I\!\!P}$}
\newcommand{\gap}{\stackrel{>}{\sim}}
\newcommand{\lap}{\stackrel{<}{\sim}}
\newcommand{\fem}{$F_2^{em}$}
\newcommand{\tsnmp}{$\tilde{\sigma}_{NC}(e^{\mp})$}
\newcommand{\tsnm}{$\tilde{\sigma}_{NC}(e^-)$}
\newcommand{\tsnp}{$\tilde{\sigma}_{NC}(e^+)$}
\newcommand{\st}{$\star$}
\newcommand{\sst}{$\star \star$}
\newcommand{\ssst}{$\star \star \star$}
\newcommand{\sssst}{$\star \star \star \star$}
\newcommand{\tw}{\theta_W}
\newcommand{\sw}{\sin{\theta_W}}
\newcommand{\cw}{\cos{\theta_W}}
\newcommand{\sww}{\sin^2{\theta_W}}
\newcommand{\cww}{\cos^2{\theta_W}}
\newcommand{\trm}{m_{\perp}}
\newcommand{\trp}{p_{\perp}}
\newcommand{\trmm}{m_{\perp}^2}
\newcommand{\trpp}{p_{\perp}^2}
%
%
\newcommand{\sqrts}{$\sqrt{s}$}
\newcommand{\Oa}{$O(\alpha_s)$}
\newcommand{\Oaa}{$O(\alpha_s^2)$}
\newcommand{\PT}{p_{\perp}}
\newcommand{\sh}{\hat{s}}
\newcommand{\uh}{\hat{u}}
\newcommand{\ttbs}{\char'134}
\newcommand{\xpomlo}{3\times10^{-4}}
\newcommand{\xpomup}{0.05}
\newcommand{\llq}{$\alpha_s \ln{(\qsq / \Lambda_{QCD}^2)}$}
\newcommand{\llqx}{$\alpha_s \ln{(\qsq / \Lambda_{QCD}^2)} \ln{(1/x)}$}
\newcommand{\llx}{$\alpha_s \ln{(1/x)}$}
%
%
%
%
\def\ar#1#2#3   {{\em Ann. Rev. Nucl. Part. Sci.} {\bf#1} (#2) #3}
\def\epj#1#2#3  {{\em Eur. Phys. J.} {\bf#1} (#2) #3}
\def\err#1#2#3  {{\it Erratum} {\bf#1} (#2) #3}
\def\ib#1#2#3   {{\it ibid.} {\bf#1} (#2) #3}
\def\ijmp#1#2#3 {{\em Int. J. Mod. Phys.} {\bf#1} (#2) #3}
\def\jetp#1#2#3 {{\em JETP Lett.} {\bf#1} (#2) #3}
\def\mpl#1#2#3  {{\em Mod. Phys. Lett.} {\bf#1} (#2) #3}
\def\nim#1#2#3  {{\em Nucl. Instr. Meth.} {\bf#1} (#2) #3}
\def\nc#1#2#3   {{\em Nuovo Cim.} {\bf#1} (#2) #3}
\def\np#1#2#3   {{\em Nucl. Phys.} {\bf#1} (#2) #3}
\def\pl#1#2#3   {{\em Phys. Lett.} {\bf#1} (#2) #3}
\def\prep#1#2#3 {{\em Phys. Rep.} {\bf#1} (#2) #3}
\def\prev#1#2#3 {{\em Phys. Rev.} {\bf#1} (#2) #3}
\def\prl#1#2#3  {{\em Phys. Rev. Lett.} {\bf#1} (#2) #3}
\def\ptp#1#2#3  {{\em Prog. Th. Phys.} {\bf#1} (#2) #3}
\def\rmp#1#2#3  {{\em Rev. Mod. Phys.} {\bf#1} (#2) #3}
\def\rpp#1#2#3  {{\em Rep. Prog. Phys.} {\bf#1} (#2) #3}
\def\sjnp#1#2#3 {{\em Sov. J. Nucl. Phys.} {\bf#1} (#2) #3}
\def\spj#1#2#3  {{\em Sov. Phys. JEPT} {\bf#1} (#2) #3}
\def\zp#1#2#3   {{\em Zeit. Phys.} {\bf#1} (#2) #3}
%
%
\newcommand{\clearemptydoublepage}{\newpage{\pagestyle{empty}\cleardoublepage}}
\newcommand{\scaption}[1]{\caption{\protect{\footnotesize  #1}}}
\newcommand{\proc}[2]{\mbox{$ #1 \rightarrow #2 $}}
\newcommand{\average}[1]{\mbox{$ \langle #1 \rangle $}}
\newcommand{\av}[1]{\mbox{$ \langle #1 \rangle $}}


\title{Hadronic structure, low x physics and diffraction}

\author{P. Marage}

\address{Universit\'e Libre de Bruxelles - CP 230, Boulevard du Triomphe,
    B-1050 Bruxelles, Belgium \\[3pt]
    E-mail: {\tt pmarage@ulb.ac.be}}

\abstract{A review is presented of numerous recent results, particularly those
submitted to the EPS-HEP99 conference: very high $Q^2$ $ep$ interactions
and direct tests of the Standard Model, new measurements of the
structure of the proton (including high x parton distributions and
tests of QCD involving the gluon distribution), low x physics (tests of
the BFKL evolution), diffraction in DIS at HERA, hard diffraction at
the Tevatron and exclusive production of vector particles at HERA.
The focus is on hard QCD features~\dag.} 

\maketitle

\fntext{\dag}{Plenary report presented at the International Europhysics 
Conference on High Energy Physics, EPS-HEP99, Tampere, Finland, 15-21 July 1999.}


\section{Introduction.}
    \label{sect:intro}

The present review covers a very large field of research, illustrated by over 
80 papers submitted to this conference, including results form HERA, the 
Tevatron and fixed target experiments. 
After a presentation of direct tests of the Standard Model (SM) performed 
at HERA at very high \qsq,
the focus of the paper is on hard QCD features~\footnote{
By lack of time and space, numerous interesting and important
topics could not be covered by the present report, in particular 
hadron final state in DIS and diffraction~\cite{final-states}, 
leading baryon studies~\cite{leading-baryons}, spin physics~\cite{spin}.}.

Time has gone when QCD needed to be tested as the theory of strong
interactions. 
The task is now to improve our understanding of the theory, i.e. provide a 
consistent 
and detailed QCD description of fundamental features of particle physics, 
in particular the structure of hadrons and diffractive scattering, and evaluate
the validity of different approximations and calculation techniques.


\section{The proton at the \boldmath{$10^{-3}$} fm scale.}

A highlight of this conference is the presentation by the H1 and ZEUS
experiments at HERA of measurements of the proton structure for 
$Q^2 \ \gapprox\ M^2_Z$, i.e. at a scale of 
$10^{-3}$ fm~\cite{Z_highqsq_xsect,H_highqsq_xsect}.
These results were obtained from the scattering of 27.5 GeV positrons with 820
GeV protons ($\sqrt {s} = 300$ GeV, 40 \pbinv\ data taken in 1994-97) and
of 27.5 GeV electrons with 920 GeV protons ($\sqrt {s} = 320$ GeV, 
16 \pbinv\ data taken in 1998-99), both in neutral current (NC) and charged 
current (CC) interactions.
They confirm, in a widely extended kinematic domain, the validity of the 
SM~\cite{Mangano}.

\begin{figure}[htbp]
\begin{center}
 \epsfig{file=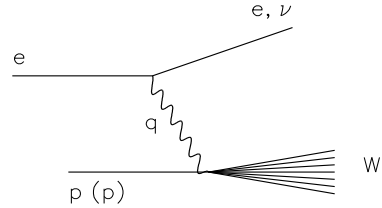,width=5.0cm,height=2.7cm}
\end{center}
\vspace{-0.3cm}
\caption{Deep inelastic $ep$ scattering.}
\label{fig:kin}
\end{figure}

The kinematics of $ep$ deep inelastic scattering (DIS) is sketched on 
Fig.~\ref{fig:kin}. The following variables are used: $Q^2 = - q^2; 
x = Q^2 / 2 p \cdot q; W^2 \simeq Q^2 / x ; y \simeq Q^2 / x \cdot s.$ 
The variable $x$ is, in the Breit frame, the fraction of the proton momentum 
carried by the struck quark;
$W$ is the invariant mass of the hadronic system; $\sqrt {s}$ is the $ep$ centre of mass
system (cms) energy.

In the SM, the cross section for DIS $ep$ scattering is given in terms of the
$F_2$, $F_L$  and $F_3$ structure functions in the following forms:
\begin{itemize}
\item {\bf {\it NC}}
\end{itemize}
\begin{eqnarray}
  \frac{ {\rm d}^2 \sigma^{NC}} {{\rm d}x \ {\rm d}Q^2 } =
  \frac {2 \pi \alpha^2} {x \ Q^4} \
  [Y_+ \ F_2(x,Q^2)                                      \nonumber \\
       - y^2 \ F_L(x,Q^2) \ _+^- \ Y_- \ x F_3(x,Q^2)],
                                                    \label{eq:xsecNC}
\end{eqnarray}
where $\alpha$ is the fine structure coupling constant and 
$Y_{\pm} = 1 \pm (1-y)^2$; 
the $-$ sign in front of the electroweak contribution proportional to $x F_3$ 
is for \epl\ scattering and the $+$ sign for \emi.

\begin{itemize}
\item {\bf {\it CC}}
\end{itemize}
\begin{eqnarray}
  \frac{ {\rm d}^2 \sigma^{CC}} {{\rm d}x \ {\rm d}Q^2 } =
  \frac {G_F^2} {4 \pi x} \ (\frac{M_W^2}{M_W^2 + Q^2})^2 \
  [Y_+ \ F_2(x,Q^2)                                       \nonumber \\
        - y^2 \ F_L(x,Q^2) \ _+^- \ Y_- \ x F_3(x,Q^2)].
                                                    \label{eq:xsecCC}
\end{eqnarray}
It is useful to get rid of the trivial $x$ and \qsq\ dependences in relations 
(\ref{eq:xsecNC}) and (\ref{eq:xsecCC}), and to define ``reduced'' cross 
sections~$\tilde \sigma$ corresponding to the quantities between brackets 
(see~\cite{Z_highqsq_xsect,H_highqsq_xsect}).

\begin{figure}[htbp]
\begin{center}
 \epsfig{file=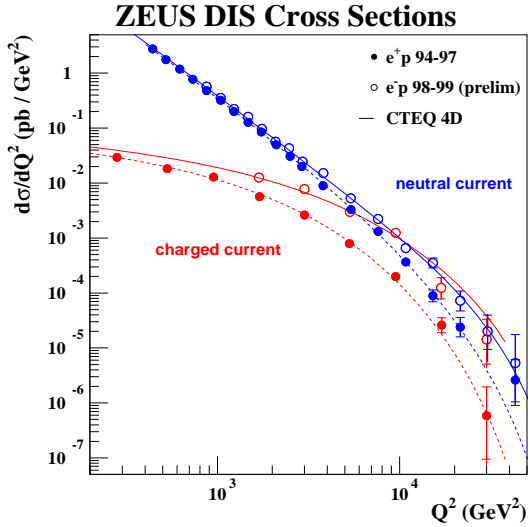,width=7.cm,height=7.cm}
\end{center}
\vspace{-0.3cm}
\caption{ZEUS measurements of \eplp\ and \emip\ NC and CC \xsec\ measurements,
  as a function of $Q^2$~\protect\cite{Z_highqsq_xsect}.
  The lines represent the SM predictions using the CTEQ4D parton distribution
  functions (pdf's).}
\label{fig:xsect}
\end{figure}

Fig.~\ref{fig:xsect} presents measurements of the \eplp\ and \emip\ NC and 
CC \xsecs~\cite{Z_highqsq_xsect}.
The similarity of the NC and CC \xsecs\ for $Q^2 \approx M^2_Z$ demonstrates
electroweak unification in the $t$ channel.

\begin{figure}[htbp]
\begin{center}
 \epsfig{file=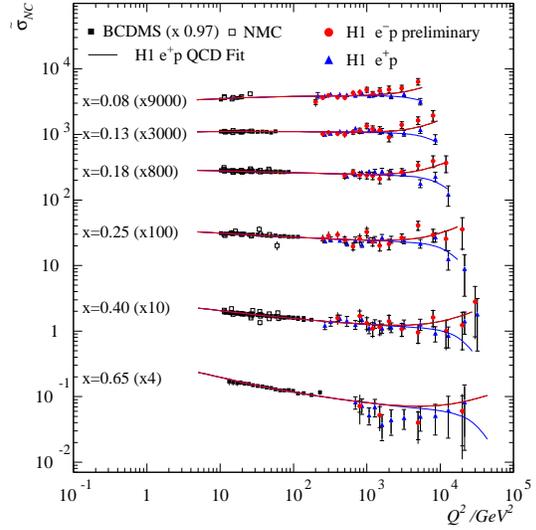,width=7.8cm,height=7.8cm}
\end{center}
\vspace{-0.3cm}
\caption{H1 measurement of \protect\eplp\ and \protect\emip\ NC cross sections,
exhibiting the effects of parity violation at high \qsq\ 
values~\protect\cite{H_highqsq_xsect}.}
\label{fig:interf}
\end{figure}

Parity violation effects due to the electroweak contribution ($\gamma - Z^{\rm o}$
interference) and corresponding to the change of sign in 
relation~(\ref{eq:xsecNC}) are visible from the difference between the \eplp\ and 
\emip\ NC \xsecs\ at high \qsq\ (see Fig.~\ref{fig:interf}; 
the effect of the small difference in $\sqrt{s}$ is negligible).
 
\begin{figure}[htbp]
\begin{center}
 \epsfig{file=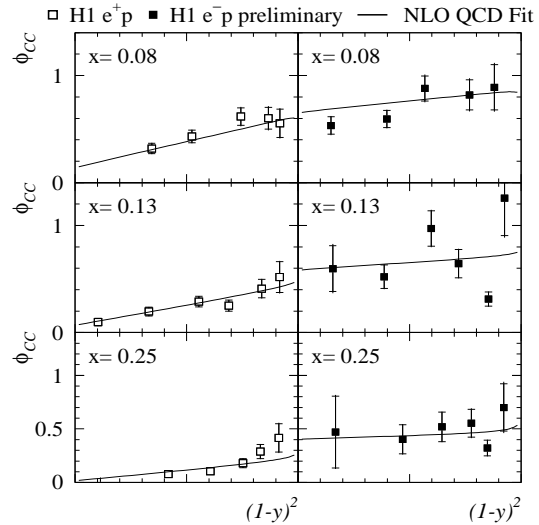,%
 bbllx=43.pt,bblly=158.pt,bburx=537.pt,bbury=680.pt,%
 width=7.cm,height=7.cm}
\end{center}
\vspace{-0.3cm}
\caption{H1 measurements of the $y$ dependence of the \eplp\ and \emip\ CC 
  reduced \xsecs~\protect\cite{H_highqsq_xsect}.}
  \label{fig:helicity}
\end{figure}

\begin{figure}[htbp]
\vspace{-0.cm}
\begin{center}
 \hspace{-0.3cm}
 \epsfig{file=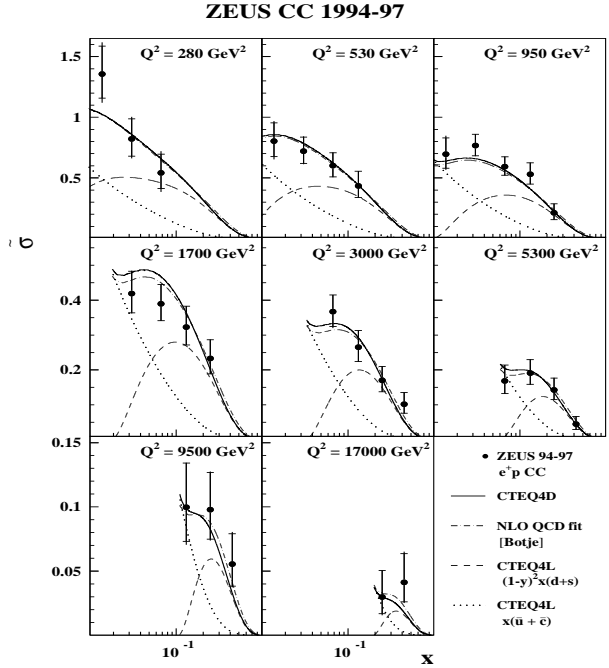,width=8cm,height=9.cm}
\end{center}
\vspace{-0.3cm}
\caption{ZEUS measurements of the \eplp\ reduced cross section $\tilde \sigma$ 
as a function of $x$ in bins of \qsq~\protect\cite{Z_highqsq_xsect}. 
The solid lines represent the SM predictions using the CTEQ4D pdf's.
The dashed and dotted lines represent the $d + s$ and $\bar u + \bar c$ 
contributions, respectively (CTEQ4L pdf's).}
  \label{fig:d-pdf}
\end{figure}

The helicity structure of the interaction is directly visible from the $y$
dependence of the CC cross sections, shown in Fig.~\ref{fig:helicity}.
The cross section for CC \emip\ \inters\ is proportional to 
$[u + c] + (1 - y)^2 [\bar d + \bar s]$, where $q$ represents the density
distribution of quark $q$ in the proton.
It is dominated by $u$ quarks, and is thus large and weakly dependent on 
$(1 - y)^2$.
In contrast, the CC \eplp\ cross section is proportional to 
$[\bar u + \bar c] + (1 - y)^2 [d + s]$, thus proportional to $(1 - y)^2$ with
a small intercept.
The contribution of $d$ quarks at high $x$ can be seen in 
Fig.~\ref{fig:d-pdf}, which presents the \eplp\ CC \xsec\ measurement as a
function of $x$ in bins of \qsq.

In conclusion, HERA has reached the space-like $Q^2 \simeq\ M_Z^2$ region with 
a measurement at the 20\% precision level of the \epl\ and \emi\ CC and NC 
\xsecs. 
This allows the direct observation of the electroweak unification, of parity
violation effects in NC and of the quark helicity structure.
 
\section{The structure of the proton.}

Understanding the structure of hadrons, in particular the proton,
is one of the main goals of particle physics but the task is difficult 
since it implies the description of long distance, non-perturbative effects.
Fortunately, factorisation applies in DIS between parton distribution 
functions (pdf's) in the target hadron (proton) and hard processes involving  
short distance interactions of partons.

The parameterisation of the pdf's and the study of their evolution according to 
the interaction scale provide information both on the proton structure 
and on the relevant features of QCD.
Their precise determination is also the base-line for any investigation of
new physics.

The functional form of the pdf's is not known theoretically. 
Empirical parameterisations, guided by theoretical arguments, are thus 
used. 
In order to reduce the number of free parameters, additional conditions 
are imposed, mainly constraining relations between different sea quark 
density distributions and between sea quark and gluon distributions.

Modern parameterisations of pdf's~\cite{GRV,CTEQ,MRST,Botje,H1-pdf,Z-pdf}
share common features:
\begin{itemize}
\item the use of NLO DLGAP~\cite{DGLAP} evolution equations;
\item a starting scale $Q_{\rm o}^2 = 1 - 2$ \gevsq\ (or even 
 lower~\cite{GRV}) for the QCD evolution;
\item the dynamical inclusion of heavy quarks~\cite{Thorn,ACOT}, needed
 since $Q_{\rm o}^2 < m_c^2$;
\item the use of essentially the same data sets.
\end{itemize}

Differences between parameterisations concern mainly:
\begin{itemize}
\item the choice of pdf's at the starting scale $Q_{\rm o}^2$ (different 
	functional forms and constraints);
\item details of the choice of data and cuts;
\item the choice of SM parameter values (\as);
\item details of the inclusion of heavy quarks.
\end{itemize}

It should be stressed that the errors on the fitted pdf's are not 
well known, which limits the significance of comparisons between 
theoretical predictions and data.
The difficulty in asserting errors on pdf's arises from the difficulty in 
controlling the following effects, several of which are addressed in the course 
of the present talk:
\begin{itemize}
\item the choice of experimental data (data of poor precision, conflicting
results);
\item the treatment of experimental errors in the data (correlated 
systematic errors);
\item the freedom of choice of the starting parameterisation form;
\item theoretical uncertainties (higher order effects, non DGLAP evolution, higher
twist contributions, nuclear effects).
\end{itemize}

\begin{figure}[t]
\vspace{0.6cm}
\begin{center}
 \epsfig{file=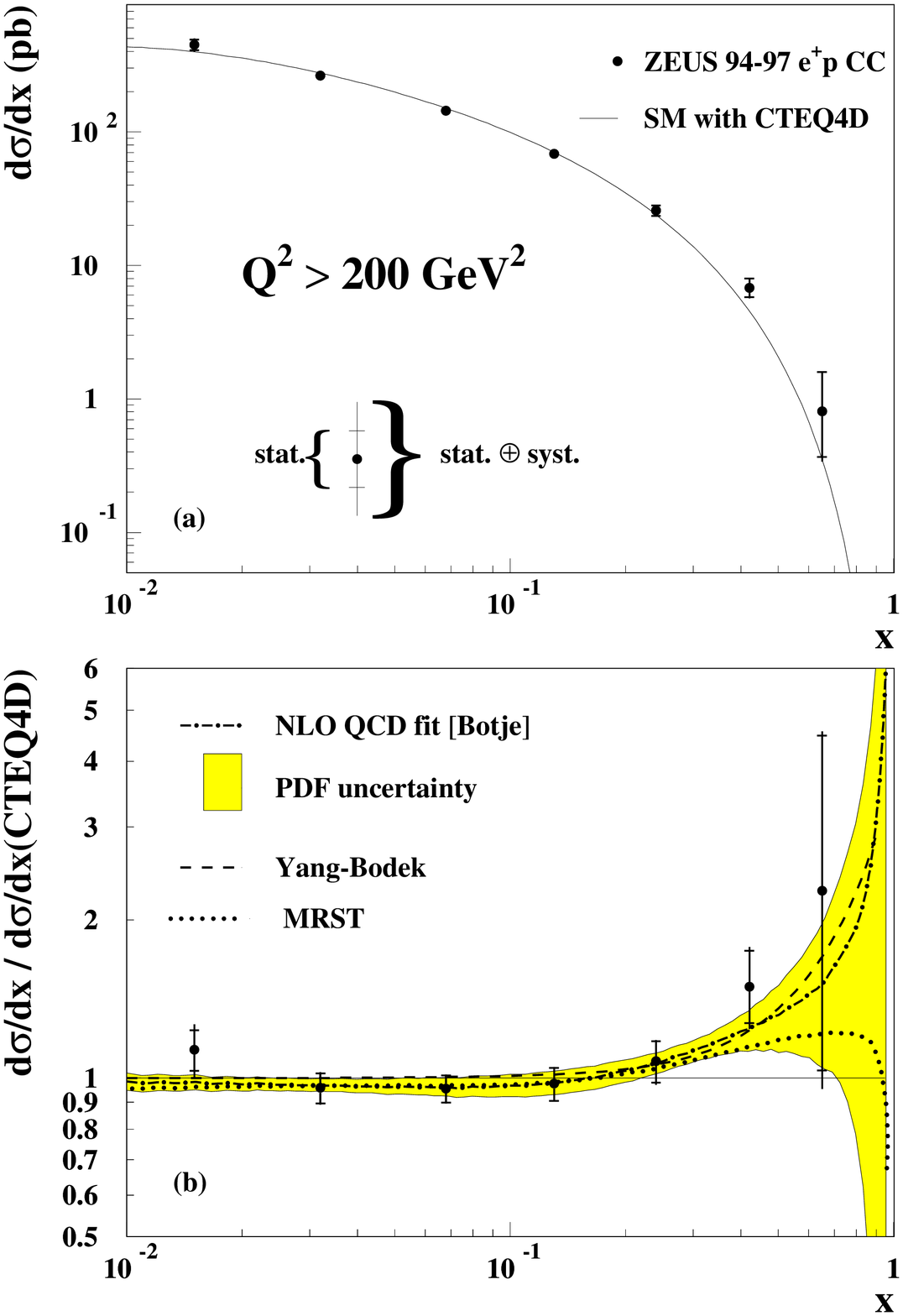,width=6cm,height=4cm}
\end{center}
\vspace{-9.1cm}
\begin{center}
 \hspace{-0.5cm}
 \epsfig{file=whitebox.eps,width=6.5cm,height=4.7cm} 
\end{center}
\vspace{4.cm}
\vspace{-0.3cm}
\caption{Ratio of the ZEUS measurement of the \eplp\ CC cross section to the SM
expectation using the CTEQ4D pdf's~\protect\cite{Z_highqsq_xsect}.
The dashed-dotted line is a NLO QCD fit~\protect\cite{Botje}; 
the associated pdf uncertainties are shown as the shaded band. 
The dashed line is the expectation for the modified $d/u$ 
ratio~\protect\cite{Yang-Bodek}.}
  \label{fig:duncert}
\end{figure}

In the low and intermediate $x$ regions, the quark distributions are well known
thanks to DIS and Drell-Yan measurements;
the precision is lower for the gluon distribution since gluons are not directly 
probed in DIS.
At higher $x$, the $d$ quark distribution for $x \gsim 0.5$ 
(see Fig.~\ref{fig:duncert}) and the gluon
distribution for $x \gsim 0.1$ are rather poorly known.

\subsection{High \boldmath{x} parton distributions.}

\subsubsection{The \boldmath{$d/u$} ratio.}

The measurement of the $d/u$ ratio of valence quarks at high $x$ is not only 
of theoretical interest, it is also important for the search for new physics 
features. 
In particular, jets with very large transverse energy (\et) with respect to the 
beam direction at the Tevatron are dominantly 
produced by quark interactions and small differences in the quark 
distributions can induce large effects on the extracted gluon density.

At high $x$, the $u$ quark distribution is well constrained by DIS on protons 
(in particular
the fixed target experiments NMC and BCDMS), but the $d$ quark distribution
is extracted from deuterium data, where Fermi motion and nuclear binding have to 
be taken into account, leading to large uncertainties.


\begin{figure}[htbp]
\vspace{-0.cm}
\begin{center}
 \epsfig{file=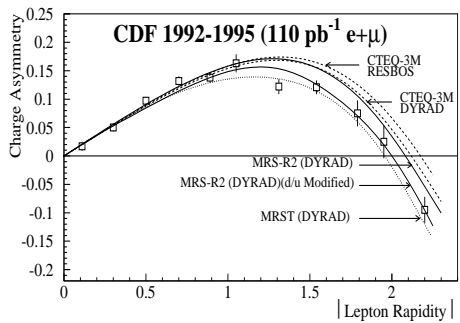,%
 bbllx=10pt,bblly=10pt,bburx=630pt,bbury=380pt,%
 width=6.cm,height=4.2cm}
\end{center}
\vspace{-0.3cm}
\caption{Rapidity distribution of the charged lepton in $W$ leptonic decay,
  measured by the CDF collaboration at the Tevatron~\protect\cite{CDF-ch-asym}.
  The effect of the modification of the $d/u$ ratio 
  limit~\protect\cite{Yang-Bodek} is shown by the difference between the 
  MRS-R2 predictions; the MRST pdf's include the present measurement.} 
  \label{fig:cdfchasym}
\end{figure}

A recent reanalysis of NMC and SLAC data~\cite{Yang-Bodek} favours
a ratio $d / u \rightarrow 0.2$ for
$x \rightarrow 1$, instead of the limit 0 which is usually chosen, albeit 
without strong theoretical motivation.
This reanalysis appears to improve the description of $\nu$-Fe \xsec,  
of jet \et\ distributions at the Tevatron, of \eplp\ CC interactions 
at HERA (although, in view of the large experimental errors,
global fits of parton distributions show little sensitivity to this 
modification of the $d / u$ ratio limit - see Fig.~\ref{fig:duncert}) and of 
the $W \rightarrow l \nu$ charge asymmetry (Fig.~\ref{fig:cdfchasym}). 

A significant improvement of the knowledge of the $d$ distribution will be obtained
from \eplp\ CC \inters\ at HERA, where no nuclear binding effects are present,
after the accelerator upgrade of year 2000 which will result in an increase by a 
factor 15 of the presently accumulated luminosity.


\subsubsection{The gluon density.}
 
\begin{figure}[htbp]
\vspace{0.5cm}
\begin{center}
  \epsfig{file=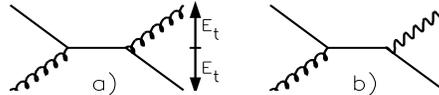,width=6cm,height=1.4cm}
\end{center}
\vspace{-0.2cm}
\caption{Two processes testing the gluon content of the proton in 
  $ p (\bar p) p$
  interactions: a) high \protect\et\ jet production ; b) prompt photon production.}
  \label{fig:gluon-diag}
\end{figure}

The main reactions relevant for the measurement of the gluon momentum distribution 
$xG(x)$ for $x > 0.1$ are high \et\ jet~(Fig.~\ref{fig:gluon-diag}a) 
and prompt photon production~(Fig.~\ref{fig:gluon-diag}b) in 
$ p (\bar p) p$ interactions.
Unfortunately, both suffer of severe problems. 

\begin{figure}[htbp]
\begin{center}
  \epsfig{file=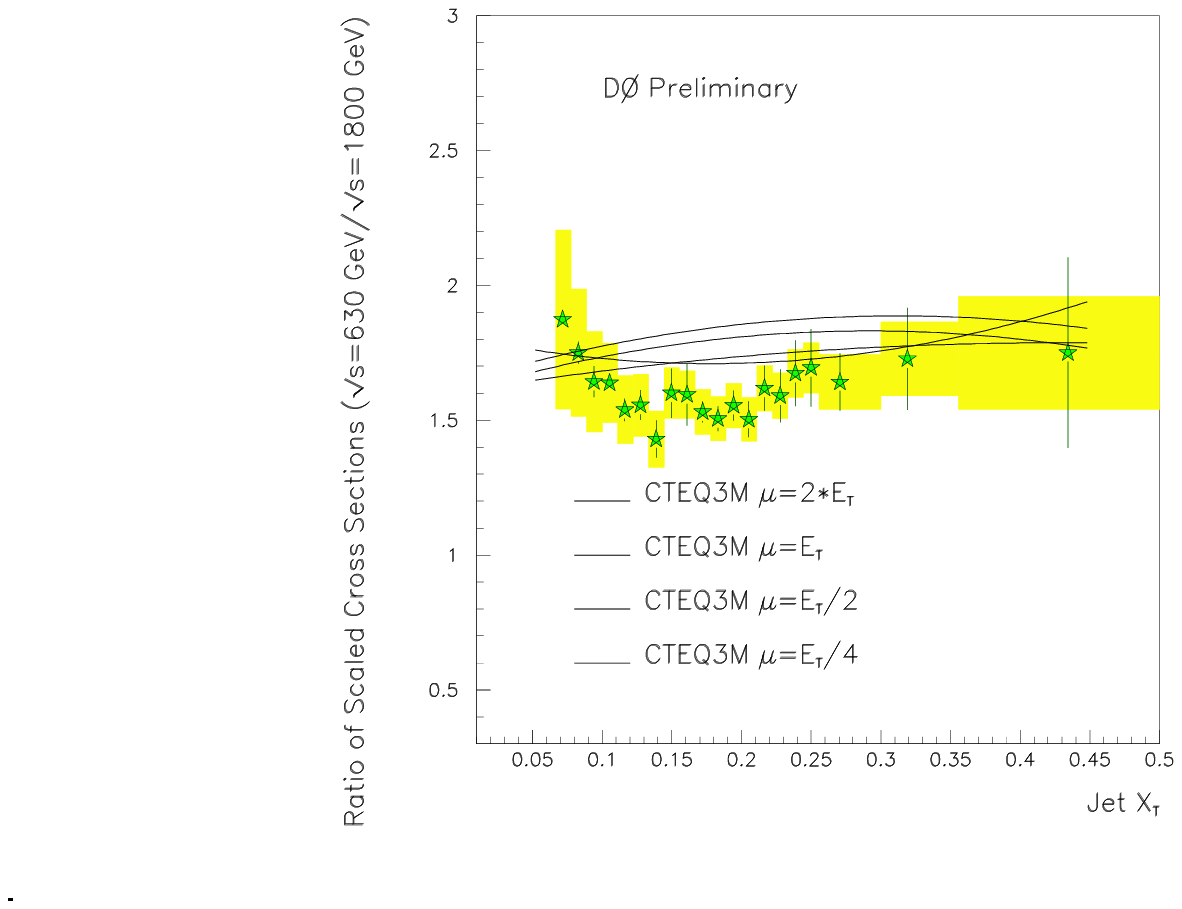,width=6.0cm,height=6.0cm}
\end{center}
\vspace{-0.3cm}
\caption{Ratio of scaled cross section for D0 jet production at 
  $\protect\sqrt{s}= 630$ GeV and $\protect\sqrt{s}= 1800$ 
  GeV, as a function of 
  $x_T = 2 E_T / \protect\sqrt{s}$~\protect\cite{D0-jet-comp}.
  The shaded area corresponds to the systematic uncertainties.
  The lines correspond to NLO calculations for different values of the QCD
  scale $\mu$.}
  \label{fig:D0-jets}
\end{figure}

High \et\ jet production has been a much debated question~\cite{Mangano}.
It now appears that D0 and CDF results are compatible within systematic
errors (including normalisation uncertainties). 
However, the D0 jet analysis reveals an inconsistency between the ratio of the
measurements at $\sqrt{s}= 630$ GeV and $\sqrt{s}= 1800$ GeV and that of the 
corresponding NLO calculations (Fig.~\ref{fig:D0-jets}). 
It is unclear whether this is an experimental problem or if it is due to a large 
influence of NNLO corrections resulting in an effective change of scale.
In addition, the extraction of the gluon distribution from high \et\ jet 
measurements is affected by the uncertainty of the $d/u$ ratio at large $x$.
In summary, the uncertainty on the gluon distribution extracted from large \et\
jets has not significantly decreased recently.

\begin{figure}[tbp]
\begin{center}
  \epsfig{file=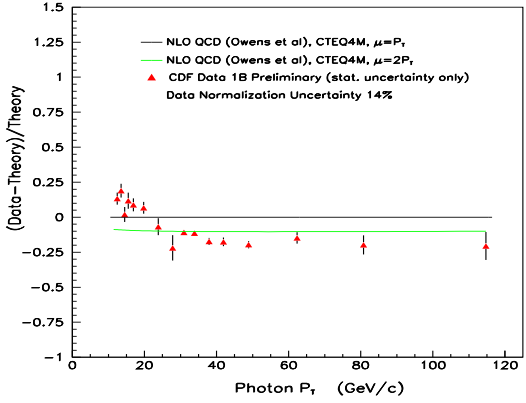,width=7cm,height=5.5cm}
\end{center}
\vspace{-0.3cm}
\caption{CDF measurement of the $p_T$ distribution of prompt photons in 
 $p \bar p$
 interactions, compared to the predictions of the CTEQ4M pdf's for two 
 different values of the QCD scale 
 $\mu$~\protect\cite{CDF-prompt-photons}.}
 \label{fig:CDF-prompt-photons}
\end{figure}

Prompt photon production is another process directly testing the gluon
content of the proton, but complications arise from the need to resum soft
gluon emission, leading to a modification of the NLO predictions.
This is parameterised in the form of an intrinsic $k_T$ contribution to the 
gluon distribution, with
$\langle k_T \rangle \simeq 1.2$ GeV for high energy fixed target data (prompt
photon and $\mu \mu$, $\gamma \gamma$, $\pi^{\rm o}$ and jet data from the E706
experiment~\cite{E706}).
At the Tevatron collider, an intrinsic $\langle k_T \rangle \simeq 3.5$ GeV 
is required to describe the prompt photon measurement by 
CDF~\cite{CDF-prompt-photons} (Fig.~\ref{fig:CDF-prompt-photons}).

Because of these large NNLO corrections, which seem not to be well under control, 
prompt photon data are not used by the CTEQ group~\cite{CTEQ}, and the gluon 
density is extracted from large \et\ jet data.
Conversely, the choice of the MRST group~\cite{MRST} is to use WA70 prompt photon 
data, with a spread of values of $\langle k_T \rangle$, and not to use 
the jet data.
In this case, different choices of $\langle k_T \rangle$ lead to significant 
differences for the absolute dijet rate predictions, but not for the shape of the 
distributions.
 
\begin{figure}[htbp]
\begin{center}
 \epsfig{file=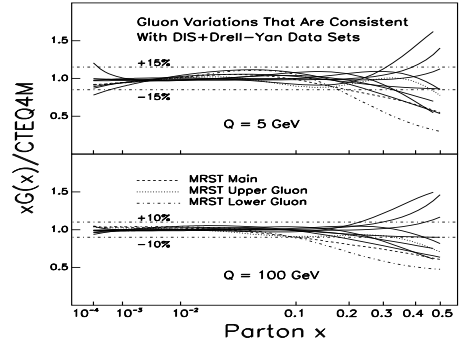,%
 bbllx=16pt,bblly=6pt,bburx=515pt,bbury=515pt,%
 width=6.0cm,height=4.5cm}
\end{center}
\vspace{-0.3cm}
\caption{Various choices of the CTEQ5 and MRST gluon momentum distributions, 
  normalised to
  the CTEQ4M parameterisation~\protect\cite{KuhlmannDIS99}.}
 \label{fig:high-x-gluon-param}
\end{figure}

In conclusion, gluon parameterisations 
can largely differ for $x > 0.1$ 
(see Fig.~\ref{fig:high-x-gluon-param}).
With increasing \qsq, the gluon density at large $x$ rapidly decreases, but the
discrepancies remain important, which has some influence
for $0.01 < x < 0.1$ because of the constraint imposed by momentum sum rules.

\subsubsection{The \boldmath{$\bar d / \bar u$} sea for 
                              \boldmath{$0.02 < x < 0.3$}.}

The Gottfried sum rule~\cite{Gottfried}, related to quark counting, states that
$$ \int_0^1 dx/x \ [F_2^p(x) - F_2^n(x)] = 1/3 $$
if $\bar u (x) = \bar d (x)$.
This is expected in perturbative QCD (pQCD), in view of the equal coupling of the
gluons to $u \bar u$ and $d \bar d$ pairs.

\begin{figure}[htbp]
\begin{center}
  \epsfig{file=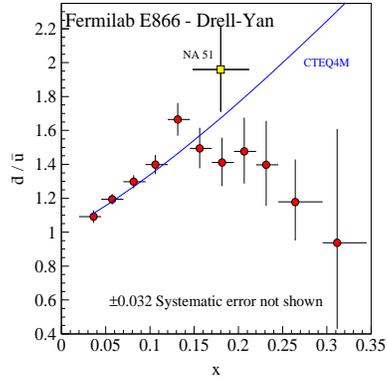,width=5cm,height=5cm}
\end{center}
\vspace{-0.3cm}
\caption{The ratio $\bar d / \bar u$ in the proton as a function of $x$
  measured by the E866 experiment~\protect\cite{E866}, compared to the CTEQ4M 
  pdf prediction, which used the NA51 data point~\protect\cite{NA51-GSR}.}
  \label{fig:E866}
\end{figure}

The NMC~\cite{NMC-GSR} and NA51~\cite{NA51-GSR} experiments have reported a
breaking of this hypothesis, respectively for $\int (\bar d - \bar u) \ dx $ 
(at $Q^2 = 4$ \gevsq) and for $x = 0.18$.
At this conference, the E866 collaboration has reported final results from
Drell-Yan proton nucleon scattering~\cite{E866}, showing that the $\bar d$ 
and $\bar u$ distributions differ for $0.02 < x < 0.3$
(see Fig.~\ref{fig:E866}). 
This is confirmed by the HERMES experiment studying charged pion 
production in $ep$ and $en$ scattering, assuming isospin 
symmetry~\cite{HERMES-GSR}.

The $\bar d / \bar u$ asymmetry is non-perturbative in origin. 
Only a small fraction of the effect is due to Pauli blocking, the main 
contribution being attributed to an asymmetry in the pion clouds accompanying 
the nucleons~\cite{Szczurek}.


\subsection{Parton distributions and QCD.}

\subsubsection{Structure functions and scaling violations.}

\begin{figure}[htbp]
\begin{center}
  \epsfig{file=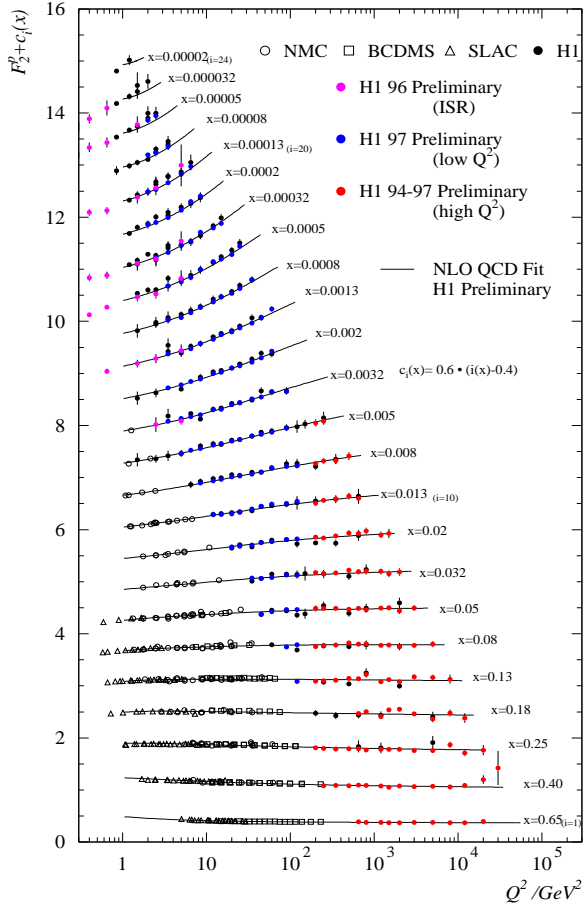,%
  bbllx=55.pt,bblly=30.pt,bburx=515.pt,bbury=770.pt,%
  width=7.8cm,height=12.cm}
\end{center}
\vspace{-0.3cm}
\caption{Measurements of the $F_2$ structure function by the H1 and fixed target
 collaborations; the lines are results of a global NLO QCD fit.}
 \label{fig:scal_viol}
\end{figure}

As shown in Fig.~\ref{fig:scal_viol}, the DGLAP QCD evolution describes $ep$ DIS
data at HERA with an impressive precision for $2 \cdot 10^{-5} < x < 0.65$ and
$1 < Q^2 < 3 \cdot 10^4$ \gevsq~\cite{H1-F2,Z-F2}. 
No need is found for higher twist or other non-DGLAP effects.

\begin{figure}[htbp]
\vspace{-0.cm}
\begin{center}
  \epsfig{file=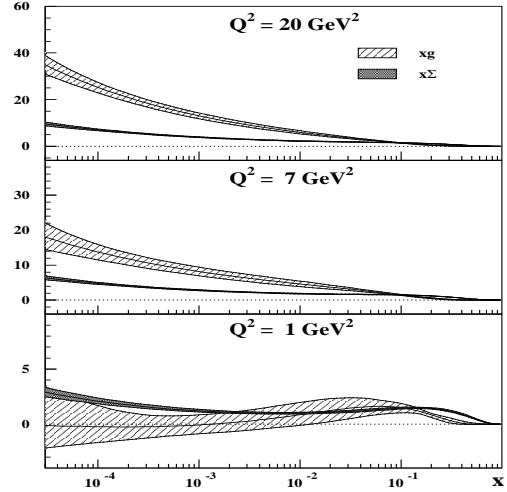,%
   bbllx=112.pt,bblly=178.pt,bburx=470.pt,bbury=650.pt,%
   width=6.5cm,height=6.5cm}
\end{center}
\caption{The gluon $xG(x)$ and the quark singlet $x \Sigma (x)$ momentum
 distributions plotted as a function of $x$ for several values of \qsq, 
 obtained from a NLO QCD fit to the ZEUS \eplp\
 cross section measurement~\protect\cite{Z-pdf}.}
 \label{fig:Z_gluon}
\end{figure}

In most of this wide kinematic domain, the $u$ and $d$ quark densities in the 
nucleon are thus precisely known. 
The gluon density distribution is not directly tested, but is extracted from
scaling violations with a good precision (see Fig.~\ref{fig:Z_gluon}).
It is successfully tested in several processes, in particular jet and charm
production.

\subsubsection{Gluons and jets.}

\begin{figure}[htbp]
\begin{center}
  \epsfig{file=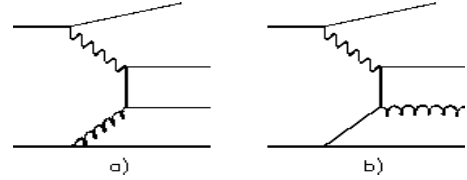,width=6cm,height=2.3cm}
\end{center}
\vspace{-0.3cm}
\caption{High \et\ jet production in DIS: a) photon gluon fusion; b) the
  QCD-Compton mechanism.}
  \label{fig:jets_diag}
\end{figure}

In DIS, high \et\ jets are mainly due to the photon gluon fusion process
(Fig.\ref{fig:jets_diag}a), with a smaller contribution from the QCD-Compton
mechanism (Fig.\ref{fig:jets_diag}b).

The differential distributions for dijet production measured by H1 and ZEUS
are in agreement~\cite{carli,terron} with predictions using 
the gluon momentum distribution extracted from scaling violations, for 
$0.005 < \xi < 0.3$ and $Q^2 > p_T^2$~\footnote{
For $Q^2 < p_T^2$, a resolved photon component may also have
to be taken into account~\cite{virtualphotonstruct}.}.
Here, $\xi$ is the fraction of the proton momentum carried by the gluon entering
the hard interaction: $ \xi = x \ (1 + M_{jj}^2 / Q^2)$, where $M_{jj}$ is the
two-jet invariant mass and $x$ the Bjorken scaling variable.
The measurement of the production rate allows a precision extraction of \as.

\begin{figure}[htbp]
\vspace{-0.cm}
\begin{center}
\epsfig{file=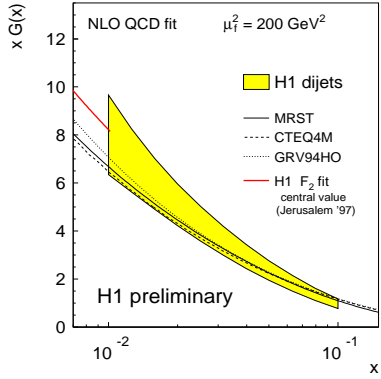,width=5cm,height=5cm}
\end{center}
\vspace{-0.3cm}
\caption{Gluon momentum distribution extracted by H1 from dijet production 
 (shaded area)~\protect\cite{carli}, compared to standard pdf's and to the 
 distribution obtained by H1 from a DGLAP fit to the inclusive cross 
 section measurement~\protect\cite{H1-pdf}.}
  \label{fig:H1-dijet-gluon}
\end{figure}
 
Conversely, using the \as\ measurement taken from other processes, a joint fit 
to the \qsq\ evolution of $F_2$ (which fixes the quark densities) and to the 
dijet rate (which drives the gluon density) can be performed.
The gluon density extracted from the dijet production~\cite{carli} is in 
agreement with that obtained from scaling violations alone 
(see Fig.~\ref{fig:H1-dijet-gluon}).


\subsubsection{Gluons and charm.}

\begin{figure}[htbp]
\begin{center}
  \epsfig{file=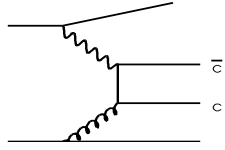,width=3cm,height=2.cm}
\end{center}
\vspace{-0.2cm}
  \caption{Charm production in DIS (photon gluon fusion process).}
  \label{fig:charm_diag}
\end{figure}

Charm production is also directly related to the gluon density, since
charm quarks are radiatively produced through the photon gluon fusion 
process (see Fig.~\ref{fig:charm_diag}).
The charm contribution to the DIS cross section, expressed in the form of a
``charm structure function'' $F_2^c$, is studied through the
decay chain 
$D^* \rightarrow D^{\rm o} \pi; D^{\rm o} \rightarrow K \pi$ or 
$K 3 \pi$~\cite{Z-F2c,H1-charm}.

\begin{figure}[htbp]
\vspace{-0.cm}
\begin{center}
  \epsfig{file=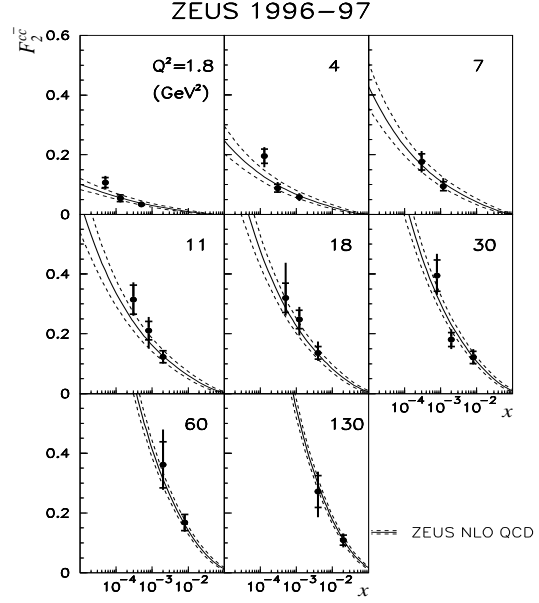,width=7cm,height=8cm}
\end{center}
\vspace{-0.5cm}
  \caption{Charm structure function $F_2^c$ measured by ZEUS as a function of $x$
  for several values of \qsq~\protect\cite{Z-F2c}.
  The curves correspond to a NLO calculation using the pdf's extracted by
  ZEUS from a QCD fit to the inclusive DIS measurement~\protect\cite{Z-pdf}.}
  \label{fig:F2c}
\end{figure}

The fast increase of $F_2^c$ with decreasing $x$ (see Fig.~\ref{fig:F2c})
confirms the gluonic origin of charm.
This increase is faster than for $F_2$, and at low $x$ (i.e. high energy $W$) 
and high \qsq, charm production accounts for some 25 \% of the DIS cross 
section~\cite{Z-F2c}.
The charm measurements by H1 and ZEUS agree well with predictions based on gluon 
momentum distributions obtained from global fits to the $F_2$ scaling 
violations.

\begin{figure}[htbp]
\vspace{-0.cm}
\begin{center}
 \epsfig{file=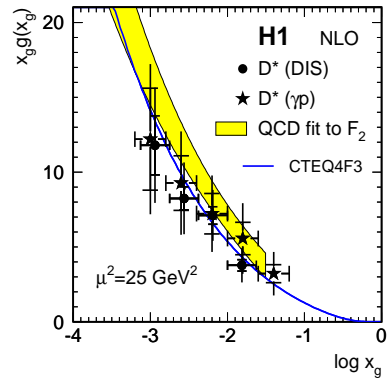,width=5cm,height=5cm}
\end{center}
\vspace{-0.3cm}
\caption{Gluon momentum distribution obtained by H1 from measurements of charm 
  production~\protect\cite{H1-charm}.
  The shaded area corresponds to a NLO QCD fit to the inclusive DIS 
  measurement~\protect\cite{H1-pdf}.
  The curve represents the CTEQ4F3 parameterisation.}
  \label{fig:H1-charm}
\end{figure}

An interesting feature of a measurement of the gluon density obtained from charm 
production (Fig.~\ref{fig:H1-charm}) is that it does not depend on a form 
assumed {\it a priori} for $x G(x)$.
However the charm measurement suffers of rather large systematic errors since 
models are needed to correct for experimental cuts and extract 
the full $D^*$ rate from the observed signal, and to 
relate the $D^*$ distributions to the charm quark distribution (effects of the
charm quark fragmentation and of final state interactions between charm quarks 
and proton remnant, leading to a beam drag).
Uncertainties also arise from the choice of the value of $m_c$.

\subsubsection{Determination of \boldmath{$F_L$}.}

\begin{figure}[htbp]
\begin{center}
 \epsfig{file=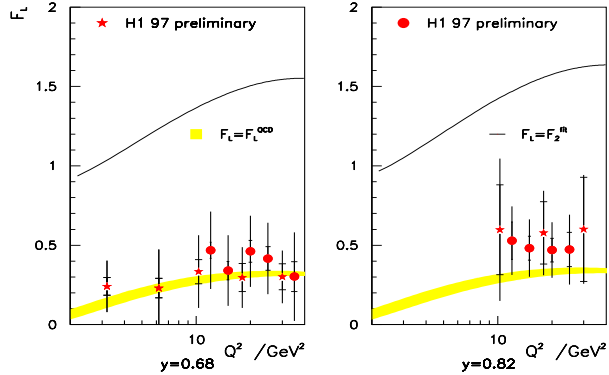,%
 bbllx=24.pt,bblly=220.pt,bburx=525.pt,bbury=583.pt,%
 width=8.cm,height=5.0cm}
\end{center}
\caption{$F_L$ determination by H1 for two values of $y$, using the subtraction
 method (dots) and the derivative method (stars)~\protect\cite{F-L}. 
 The grey areas represent the QCD predictions; the lines represent the case 
 where $F_L = F_2$.}
 \label{fig:F_L}
\end{figure}

Following relation~(\ref{eq:xsecNC}), the differential NC cross section is
proportional (for $Q^2 \ll m_Z^2$) to the reduced cross section
$\sigma_r = F_2(x,Q^2) - y^2 / Y_+ \ F_L(x,Q^2).$
Two consistent determinations of the longitudinal structure function $F_L$
have been obtained by H1 at large $y$~\cite{F-L} (Fig.~\ref{fig:F_L}):
\begin{itemize}
\item the QCD evolution of the pdf's is assumed to be valid at large $y$, and
$F_L$ is computed by the subtraction of the $F_2$ contribution from $\sigma_r$;
\item a linear extrapolation of the derivative 
  $\partial F_2 / \partial \log y$ is assumed for large $y$, providing a
  determination of $F_L$.
\end{itemize}
These determinations are consistent with QCD predictions, which are driven by the
gluon distribution in the proton.

\subsection{Conclusion.}

In conclusion, the proton structure function $F_2(x, Q^2)$ is measured over a 
huge kinematic 
domain, and QCD fits describe the scaling violations with high precision.
Except for uncertainties at high $x$ for the $d/u$ ratio and for the gluon
density, the parton distributions are thus precisely known. 
In particular, the gluon density extracted from fits to the scaling 
violations in the intermediate $x$ domain is in good agreement with 
measurements of dijet and charm production and with determinations of $F_L$.


\section{Low x physics.}
\begin{figure}[htbp]
\vspace{-0.cm}
\begin{center}
\epsfig{file=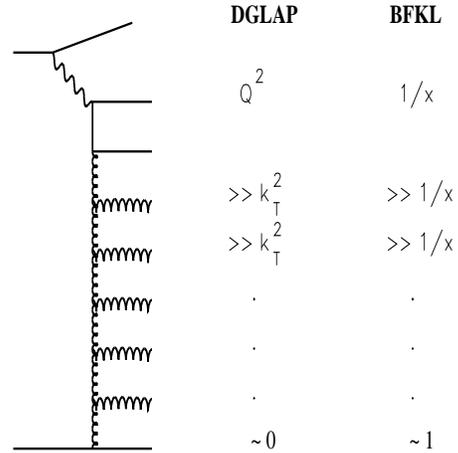,width=6cm,height=6cm}
\end{center}
\vspace{-0.2cm}
  \caption{Two limiting cases of QCD evolution in DIS: high virtuality 
  (DGLAP evolution) and high energy (BFKL evolution).}
\label{fig:2evol}
\end{figure}

In DIS, parton emission (mainly gluons) between the struck quark and the target 
remnant can be described for two limits, calculable in pQCD 
(see Fig.~\ref{fig:2evol}):
\begin{itemize}
\item the high virtuality limit (large \qsq), described by the DGLAP evolution
equations~\cite{DGLAP} which correspond to a strong ordering in $k_T$ of the
emitted gluons (from
$k_T^2 \simeq Q^2$ at the photon vertex to $k_T^2 \simeq 0$ at the target vertex),
with resummation of the $[\as \log \qsq]^n$ terms (LO). 
In this limit, $k_T$ is thus small for a large $x$ gluon.
\item the high energy limit (small $x$, with $W^2 \simeq Q^2/x$), described by the
BFKL equations~\cite{BFKL} which correspond to a strong ordering in $1/x$ (from very
small $x$ to $x \simeq 1$), with resummation of the $[\as \log 1/x]^n$ terms.
In this case, there is no $k_T$ ordering and $k_T$ can be large even for a large
$x$ gluon.
\end{itemize}

A striking prediction of the BFKL evolution at LO is a strong energy dependence of
the cross section:
$\sigma(s) \propto s^{\alpha_{BFKL} -1} \simeq s^{0.4 - 0.5}$,
whereas in ``soft'' hadron--hadron interactions~\cite{DoLa}, only a weak 
energy dependence of the cross section is observed:
$\sigma(s) \propto s^{0.08-0.10}$
(here, $s$ is the square of the total hadronic energy, denoted by $W$ in 
DIS)~\footnote{
First studies of NLO contributions~\cite{BFKL-NLO} indicated that the
corresponding corrections can be very large, suggesting an unstable behaviour of
the calculation. Recently, higher order corrections were found to be better under
control when using more ``physical'' renormalisation schemes than the 
${\overline {MS}}$ scheme~\cite{brodsky-BFKL-NLO,other-BFKL-NLO}.}.

\begin{figure}[htbp]
\vspace{-0.cm}
\begin{center}
 \epsfig{file=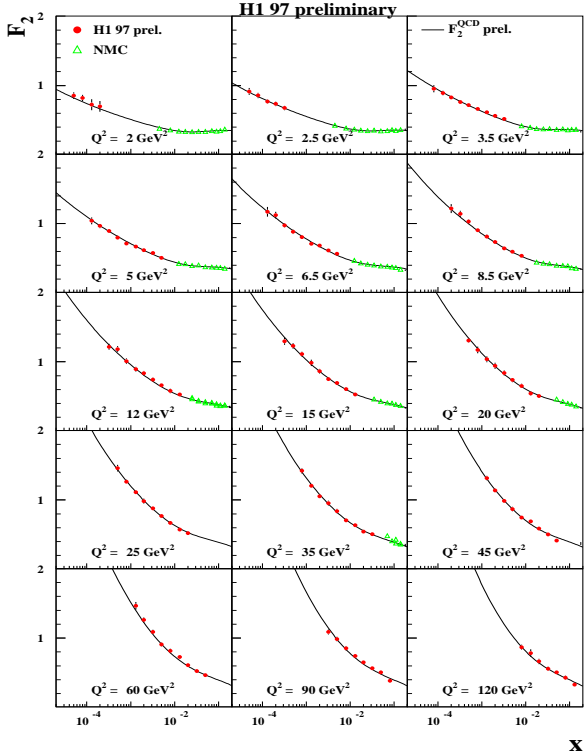,%
 bbllx=30.pt,bblly=40.pt,bburx=530.pt,bbury=777.pt,%
 width=7.8cm,height=10cm}
\end{center}
\vspace{-0.3cm}
  \caption{Measurement of the $F_2(x,Q^2)$ structure function by the H1,
  NMC and BCDMS collaborations as a function of $x$ in bins of 
  \qsq~\protect\cite{F-L}.
  The lines show the result of a NLO DGLAP fit.}
  \label{fig:q2bins}
\end{figure}

\begin{figure}[htbp]
\vspace{-0.cm}
\begin{center}
 \epsfig{file=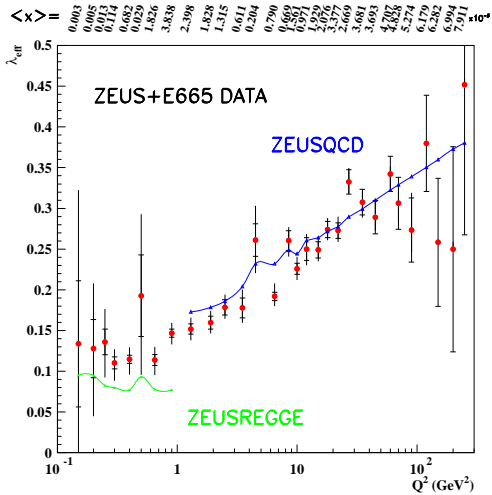,%
 bbllx=43.pt,bblly=150.pt,bburx=557.pt,bbury=662.pt,%
 width=6.5cm,height=6.5cm}
\end{center}
\vspace{-0.3cm}
\caption{ZEUS measurement of the parameter 
 $\lambda = {\rm d} \ln F_2 / {\rm d} \ln (1/x)$ describing, for fixed
 $Q^2$, the rise of $F_2$ towards low $x$~\protect\cite{Z-pdf}.
 The lines represent the ZEUS NLO DGLAP fit for 
 $Q^2 > 1$ \gevsq\ and a Regge type parameterisation for $Q^2 < 1$ \gevsq.}
  \label{fig:lambda}
\end{figure}

The most important result at HERA is probably the observation of a fast 
increase of the $F_2$ structure function at low $x$ in the DIS regime
(see Fig.~\ref{fig:q2bins}), attributed to the increase of the gluon density.
This is parameterised for $x < 0.1$ in the form 
$F_2 (x, Q^2) \propto x^{-\lambda}$ (Fig.~\ref{fig:lambda}).
Whereas at small \qsq\ $\lambda$ is low and close to the ``soft'' value 
0.08-0.10~\cite{Z-pdf,F-L}
the high value of $\lambda$ measured at high \qsq\ may be consistent with 
a BFKL interpretation of the $x$ evolution of the structure function
(remember that $1/x \propto W^2$). 
However, this behaviour is also compatible with a DGLAP-type evolution, as 
demonstrated by the quality of the DGLAP fits to the \qsq\ evolution in 
Figs.~\ref{fig:scal_viol} and~\ref{fig:q2bins}~\footnote{
Note that the freedom of choice of the pdf parameterisations at the starting
value  of the DGLAP evolution may ``hide'' BFKL features. 
Note also that gluon emissions (``rungs'' of the BFKL ladder) are separated by 
about two units in rapidity, implying that only a small number of ``rungs'' 
plays a role at HERA energies. 
The rapidity of a particle is given with respect to a given axis $z$ as 
$y = \frac {1}{2} \log \frac {E + p_z} {E - p_z}$; the rapidity interval 
between two particles is invariant under a boost along $z$.}.

The relevance of the BFKL approach can thus not be demonstrated on the basis of 
the total cross section measurements alone.
Footprints for BFKL evolution are to be searched for specifically in 
exclusive channels, in particular those characterised by both a strong energy
evolution and the absence of a strong $k_T$ ordering.
In a high energy (very low $x$) DIS process, a marked difference characterises 
the emission of partons carrying a large fraction of 
the proton momentum: for BFKL, such partons can be emitted with a
large $k_T$, whereas for DGLAP they are restricted to small $k_T$ values.


\subsection{Large energy, large $p_T$ {\rm {\it $\pi^{\rm o}$}} production 
            at HERA.}

\begin{figure}[htbp]
\vspace{-0.cm}
\begin{center}
  \epsfig{file=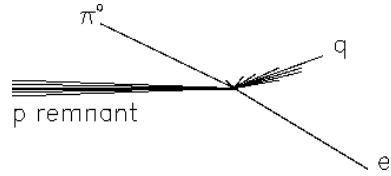,width=5cm,height=2.2cm}
\end{center}
\vspace{-0.3cm}
\caption{Final state topology for large energy, large $p_T$ 
 $\pi^{\rm o}$ emission in DIS.}
 \label{fig:pi0diag}
\end{figure}

The process $e^+ p \rightarrow e^+ \pi^{\rm o} X$ has been studied by 
H1~\cite{pi0H1} for DIS events with large $\pi^{\rm o}$ energy and large 
$p_T^{\pi^{\rm o}}$ (defined with respect to the $\gamma^{\ast} p$ axis): 
$x_{\pi^{\rm o}} = p_{\pi^{\rm o}} / p_p > 0.01$, $p_T^{\pi^{\rm o}} > 2.5$ GeV, 
for events with $Q^2 > 2$ \gevsq\ and $5 \cdot 10^{-5} < x < 5\cdot 10^{-3}$.
For such events, the photon virtuality \qsq\ and the transverse momentum squared
of the parton emitted in the parton cascade, $k_t^2$, are thus of similar 
magnitudes.
The $\pi^{\rm o}$ meson is emitted close to the proton direction 
(``forward'' direction), and is well separated in rapidity from the quark jet 
(see Fig.~\ref{fig:pi0diag})~\footnote{
A related process is the emission of a ``forward'' jet~\cite{fw_jet}.
However the acceptance in the forward direction for jet reconstruction is 
reduced compared to that for detecting a $\pi^{\rm o}$ meson.}.

\begin{figure}[htbp]
\vspace{6.cm}
\begin{center}
\begin{picture}(8.0,6.0)
  \setlength{\unitlength}{1.0cm}  
  \put(-4.,-0.5){\epsfig{file=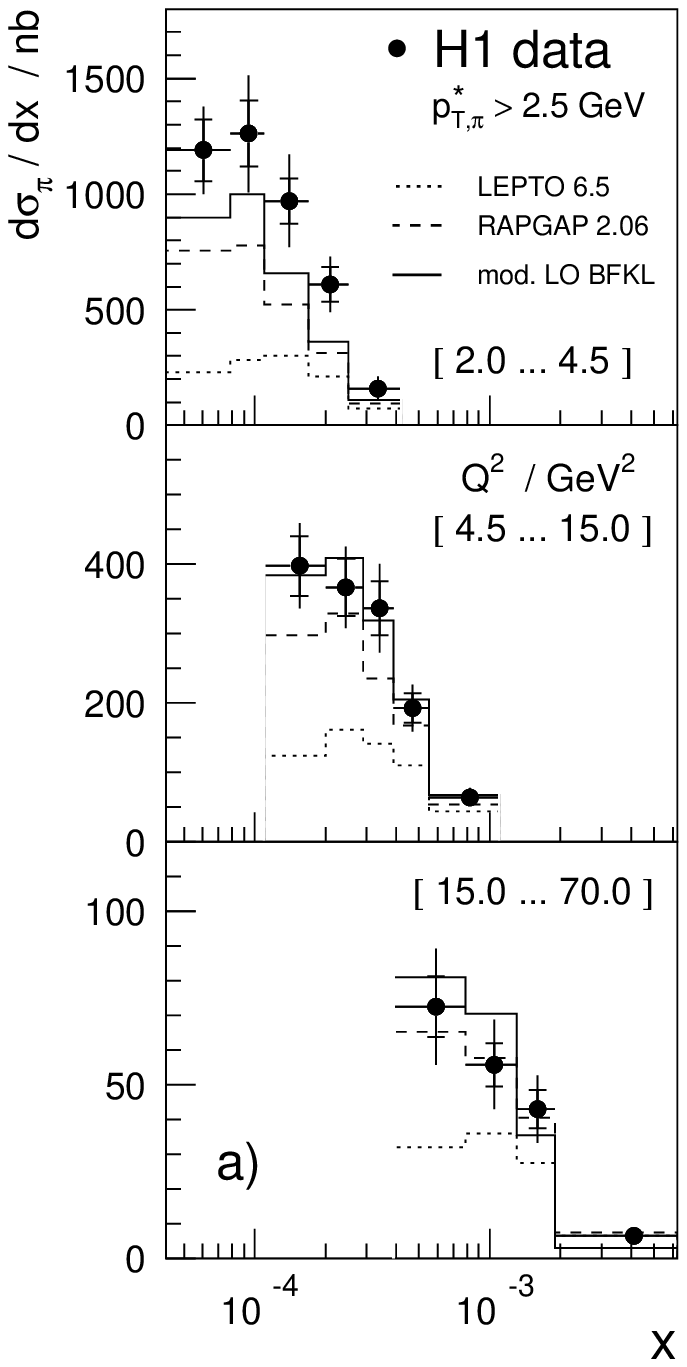,width=4.2cm,height=7cm}}
  \put(0.,-0.5){\epsfig{file=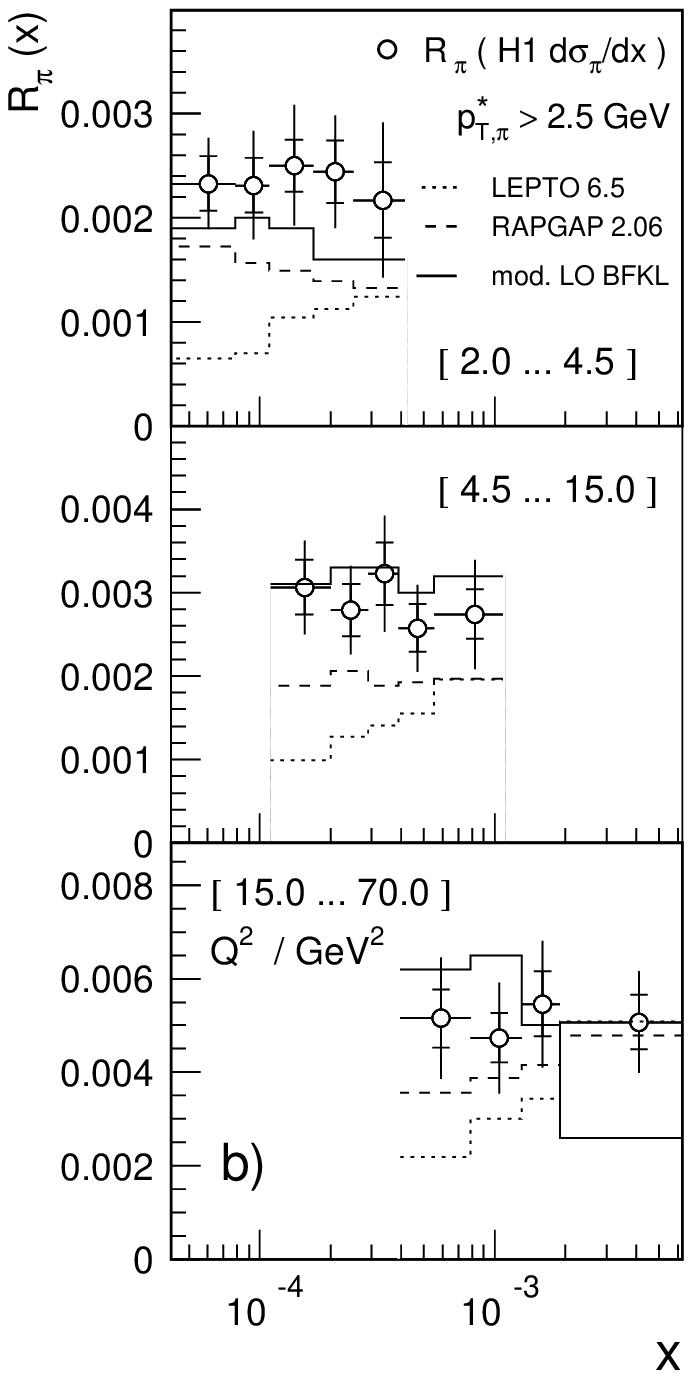,width=4.2cm,height=7cm}}
\end{picture}
\end{center}
\caption{H1 measurement of $\pi^{\rm o}$ production with 
  $p_{\pi^{\rm o}} / p_p > 0.01$, $p_T^{\pi^{\rm o}} > 2.5$ GeV,
  as a function of $x$ in 3 intervals of \qsq~\protect\cite{pi0H1}:
  left) cross section; right) production rate in DIS.
  The full histograms represent the predictions of the (modified) LO BFKL 
  model~\protect\cite{BFKLmodel}; 
  the dotted histograms are the predictions of the LEPTO6.5 
  model~\protect\cite{LEPTO6.5};
  the dashed histograms correspond to a model which includes a resolved
  photon contribution in DIS~\protect\cite{rapgag}.}
\label{fig:pi0H1}
\end{figure}

As shown in Fig.~\ref{fig:pi0H1}, the absolute cross section and the production
rate for these events are consistent with predictions of a (modified) LO BFKL 
model~\cite{BFKLmodel} for several intervals in \qsq.
They are not compatible with the predictions of the LEPTO6.5 model~\cite{LEPTO6.5}, 
which is based on the DGLAP evolution.
A model~\protect\cite{rapgag} which includes a resolved photon 
contribution in DIS~\cite{virtualphotonstruct} gives a better, but not 
satisfactory description of the data.

\subsection{Dijets with a large rapidity separation at the Tevatron.}

\begin{figure}[htbp]
\vspace{-0.cm}
\begin{center}
  \epsfig{file=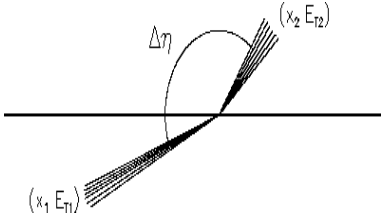,width=5cm,height=2.8cm}
\end{center}
\vspace{-0.5cm}
\caption{Two jets ($x_1, E_{T1}$) and ($x_2, E_{T2}$), separated by a large
  pseudorapidity gap $\Delta \eta$.}
\label{fig:mnjetsdiag}
\end{figure}

The production, e.g. in $p \bar{p}$ interactions, of two high $E_T$ jets 
separated by a large gap $\Delta \eta$ in 
(pseudo-)rapidity~\footnote{ 
The pseudorapidity is given by $\eta = - \ln \tan (\theta / 2)$; it corresponds 
with the rapidity in the limit of vanishing mass.} 
(see Fig.~\ref{fig:mnjetsdiag})
can also typically be described in a BFKL approach~\cite{MNjets}: the larger 
the gap in rapidity, the larger the number of ``rungs'' (gluon emissions) in the 
BFKL ladder.

The cross section for this process is given following the BFKL evolution
for jets of transverse energies $E^T_1$, $E^T_2$ by the relation:
\begin{eqnarray}
  \sigma (x_1, x_2, Q^2, \Delta \eta) \propto
    x_1 P(x_1, Q^2) \ x_2 P(x_2, Q^2)  \nonumber \\
    \frac {1} {Q^2}
    \frac {e^{(\alpha_{BFKL}-1) \ \Delta \eta}} {\sqrt {\as \ \Delta \eta}},
                                              \label{eq:MNjets}
\end{eqnarray}					      
where $x_1$ and $x_2$ are, respectively, the fractions of the beam
particle energies carried by the partons entering the strong interaction, 
$xP(x,Q^2)$ being the (colour weighted) sum of the gluon and quark distribution 
functions, and $Q^2 \simeq E^T_1 \cdot E^T_2$.

In relation~(\ref{eq:MNjets}), the test of the BFKL evolution is provided by 
the $\Delta \eta$ dependence of the cross section.
At a given beam energy, varying $\Delta \eta$ means changing $x_1$ and $x_2$,
which leads to uncertainties due to the pdf's.
For this reason, the measurement was performed by the D0
collaboration~\cite{D0MNjets} for jets with $E_T > 20$ GeV, for two different beam 
energies (with $\sqrt {s} =
630$ and $1800$ GeV, respectively) but for fixed values of $x_1$, $x_2$ and
$Q^2$, and thus different values of $\Delta \eta$.
The ratio $R$ of the two cross sections is given by
$R_{1800/630} = e^{[\alpha_{BFKL} -1] 
        [\Delta \eta _{1800} - \Delta \eta _{630}]}
	\ / \ [\Delta \eta _{1800} / \Delta \eta _{630}]^{1/2}.$

The D0 measurements gives the value  $R_{1800/630} = 2.9 \pm 0.3$ (stat.) 
$\pm 0.3$ (syst.) for $\langle \Delta \eta _{630} \rangle = 2.6$ and 
$\langle \Delta \eta _{1800} \rangle = 4.7$.
This value is incompatible with a QCD LO evolution, which asymptotically tends
to 1 as $\Delta \eta$ increases.
It is suggestive of a BFKL evolution but the present measurement would correspond 
to the high value $\alpha_{BFKL} = 1.7 \pm 0.1 \pm 0.1$.

\subsection{Conclusion.}
In summary, considerable theoretical work is providing increasingly reliable 
and stable higher order calculations of the BFKL evolution.
On the experimental side, measurements of processes characterised by large
rapidity separations between partons suggest the presence of BFKL processes.
However Monte Carlo simulations including higher
order contributions and details of hadron fragmentation are necessary
in order to provide conclusive tests of BFKL predictions.


\section{Diffraction.}

\subsection{Introduction.}

Understanding diffractive interactions is of 
fundamental importance for the understanding of elementary particle physics
since diffraction governs the high energy behaviour of elastic cross sections 
and thus of total cross sections (this relation is provided by the optical 
theorem, which derives from the unitarity of the S-matrix).

Moreover, the hypothesis of analyticity of the S-matrix and the crossing 
property of elementary particle processes allow relating the physical 
amplitudes in the $s$- and $t$-channels.
In particular, the energy dependence of total cross sections in the 
$s$-channel is related to the properties (quantum numbers) of the particle
states which can be exchanged for elastic scattering in the $t$-channel.

In the framework of Regge theory~\cite{CollinsRegge}, the concept of exchange 
of particles in the $t$ channel is extended to the exchange of ``trajectories'',
defined in the squared four-momentum / angular momentum ($t, \alpha$) plane.
The mass squared and the spin of real particles with related quantum numbers 
are observed to define linear trajectories: 
$\alpha(t) = \alpha(0) + \alpha^\prime \cdot t$.
This linear behaviour prolongates in the negative $t$, virtual exchange domain.
The energy dependence of cross sections is thus governed by the intercept 
$\alpha$ and the slope $\alpha ^{\prime}$ of the relevant trajectories.


For total cross sections, the optical theorem leads, when neglecting the real 
part of the elastic amplitudes, to the relation
$\sigma_{tot} \propto s^{\alpha(0) - 1}$.
Among known particles, the $\rho$ and $f$ meson families (``reggeon'' 
trajectory) have the highest intercept, with $\alregg(0) \simeq\ 0.5$,
implying that $\sigma \propto 1 / \sqrt{s}$ for processes mediated by 
reggeon exchange; for the pion family, 
$\alpha_{\pi}(0) \simeq\ 0$ and $\sigma \propto 1 / s$.

At high energy, the total hadron--hadron cross section is however known not 
to decrease, but to increase slightly with energy: 
$\sigma_{tot}^{hh} \propto s^{0.08-0.10}$~\cite{DoLa}.
This behaviour is thus attributed to the exchange of an object which cannot
be related to known hadrons and is found to carry the quantum numbers 
of the vacuum: the pomeron.

It is a challenge for QCD to provide a ``microscopic'' picture of the pomeron
(see e.g.~\cite{NNN,pech,BEKW,hebecker}).
The simplest model is a two-gluon system, in contrast with reggeons and other 
mesons which are fundamentally two-quark systems 
(glueballs are thus possibly physical states related to the pomeron).
Any QCD description of high energy scattering needs to account for the 
pomeron properties, in particular the increase of total cross sections with 
energy.
The observed power-law for this increase is however incompatible at very high
energy with bounds arising from the unitarity of the S-matrix (Froissart bound). 
It is thus a major task to understand how QCD offers a mechanism for the 
damping of the total cross section at high energy.

It should be stressed that alternative models aim at explaining diffraction
by soft colour recombination of partons, without a reference to the concept of 
pomeron~\cite{LEPTO6.5,eboli}.

\subsection{Diffraction in DIS at HERA.}

\subsubsection{Diffractive structure function and energy dependence.}
                                            \label{sect:endep-f2d3}

\begin{figure}[htbp]
\vspace{-0.cm}
\begin{center}
 \epsfig{file=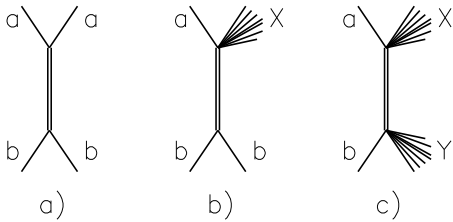,width=6cm,height=3cm}
\end{center}
\vspace{-0.5cm}
\caption{a) elastic scattering; b) single diffractive dissociation; 
 c) double diffraction.}
 \label{fig:diffr}
\end{figure}

The experimental study of the pomeron structure is facilitated by
a process which generalises elastic scattering: diffractive dissociation
$a + b \rightarrow X + b$, with $M_X \ll \sqrt{s}$, the ($ab$) cms energy 
-- see Fig.~\ref{fig:diffr}
(in ``double diffraction'', both states $a$ and $b$ are excited into small 
mass systems).
Diffractive dissociation is explained by the differential absorption by the
target of the various hadronic states which build up the incoming 
state~\cite{GoodWalker}. 

\begin{figure}[htbp]
\vspace{-0.cm}
\begin{center}
 \epsfig{file=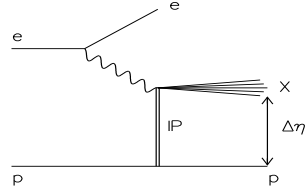,width=4.cm,height=2.5cm}
\end{center}
\vspace{-0.3cm}
\caption{Diffractive dissociation at HERA.}
\label{fig:DDIS}
\end{figure}

It was an important observation at HERA that 8 to 10 \% of the DIS cross 
section is due to diffractive dissociation (Fig.~\ref{fig:DDIS}).
These events are characterised by a large gap in (pseudo-)rapidity 
$\Delta \eta$, devoid of 
hadronic energy, between the hadronic system $X$, of mass $M_X$, and the 
scattered proton (or the baryonic system $Y$ resulting from proton excitation), 
implying the exchange of a colour singlet system.
The gap is kinematically related to a small value of $M_X$, $M_X \ll W$; 
for small \qsq, the momentum fraction lost by the proton (or the excited 
system) is $x_L \simeq\ M_X^2 / W^2 \ll 1$.

A unique tool for testing the structure of the pomeron is thus provided at
HERA by diffractive deep inelastic scattering (DDIS).
Following the model of inclusive DIS, a ``diffractive structure function'' 
$F_2^{D(3)}\,(x_{I\!\!P},\,\beta,\,Q^2)$ is extracted from the inclusive DDIS 
cross section~\cite{f2d3H1,H1diffrvan,f2d3ZEUS,f2d3ZEUSlowq2,f2d3ZEUSfps}, 
with $x_{I\!\!P} \simeq (Q^2 + M_X^2) / (Q^2 + W^2) \simeq 1 - x_L $,
$\beta \simeq Q^2 / (Q^2 + M_X^2)$ and $x = \xpom \cdot \beta$~\footnote{
When diffractive events are selected by the presence of a gap in rapidity 
devoid of hadronic energy, the four-momentum squared $t$ at the proton vertex 
is usually not measured, and the measurements are integrated over $t$. 
With the use of their proton spectrometer, the ZEUS experiment has performed a 
measurement of the $t$ distribution~\cite{tslope}.}.

It has been proven in pQCD~\cite{DDISfactor} that the amplitudes for DDIS 
processes factorise into a part which depends on $\xpom$ (a ``pomeron flux
factor''), and a ``structure function'' $\fiipomfull$ corresponding to a 
universal partonic structure of diffraction~\cite{fracture}.
The variables $\xpom$ and $\beta$ can thus be interpreted, respectively,
as the fraction of the proton momentum carried by the pomeron, and the 
fraction of the pomeron momentum carried by the struck quark.

In a Regge approach, the ``pomeron flux factor'' follows a power law:
$F_2^{D(3)}\,(x_{I\!\!P},\,\beta,\,Q^2) \propto 
(1/\xpom)^{2 \alpha_{\pom} -1} \cdot \tilde F_2^{D}\,(\beta,\,Q^2)$.

In photoproduction, HERA measurements~\cite{h1_photoprod,zeus_photoprod}
give for the pomeron intercept 
values consistent with the ``soft'' value $1.08 - 1.10$.
In DIS, the pomeron intercept 
$\alpha_{\pom}(0)$ is significantly higher~\footnote{
In the HERA energy range, pomeron exchange dominates rapidity gap events for
$\xpom \lsim 0.01$; for higher $\xpom$ values (lower energy), reggeon exchange 
has also to be taken into account (see e.g.~\cite{f2d3H1}).}: 
the H1 measurement~\cite{f2d3H1} is 
$\alpha_\pom(0) = 1.20 \pm 0.02 {\rm \ (stat.)} \pm 0.01 {\rm \ (syst.)}
                        \pm 0.03 {\rm \ (model)}$,
and the ZEUS measurement~\cite{f2d3ZEUS} is		
$\alpha_\pom(0) = 1.16 \pm 0.01 {\rm \ (stat.)} 
                     \ ^{+0.04}_{-0.01} {\rm \ (syst.)}$
($\alpha_\pom(0)$ has here been computed from the ZEUS measurement of
$\alpha_\pom(\bar t)$ using $\alpha^\prime = 0.25$ \gevsqm\ and 
$| \bar t | = 1 / 7.1$ \gevsq~\cite{tslope}).

\begin{figure}[htbp]
\vspace{-0.cm}
\begin{center}
 \epsfig{file=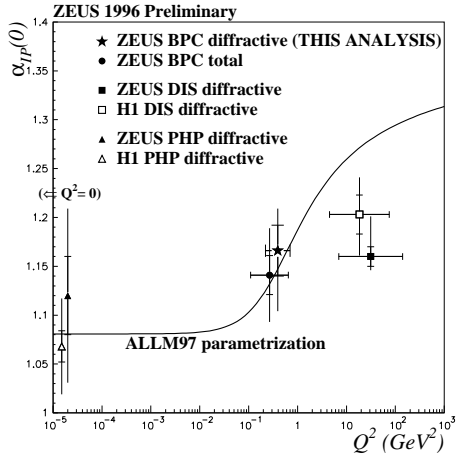,%
 bbllx=0.pt,bblly=10.pt,bburx=525.pt,bbury=535.pt,%
 width=6.cm,height=6.cm}
\end{center}
\vspace{-0.3cm}
\caption{Measurements of $\alpha_\pom(0)$ as a function of 
 \qsq~\protect\cite{f2d3ZEUSlowq2}.
 The curve represents the total $\gamma^\ast p$ cross section, in the ALLM 
 parameterisation~\protect\cite{ALLM}.}
\label{fig:z-diffr-bpc}
\end{figure}

At this conference, ZEUS has presented 
the measurement~\cite{f2d3ZEUSlowq2} 
$\alpha_\pom(0) = 1.17 \pm 0.03 {\rm \ (stat.)} 
                     \ ^{+0.04}_{-0.06} {\rm \ (syst.)}$
in the range $0.22 < Q^2 < 0.70$ \gevsq\ (Fig.~\ref{fig:z-diffr-bpc}). 
A transition from a soft to a hard behaviour thus happens at  
low \qsq\ values.
It should be noted that the value of $\alpha_\pom(0)$ extracted from the 
diffractive cross section at low \qsq\ is similar to that obtained from the
total $\gamma^\ast p$ cross section in this domain 
(see Fig.~\ref{fig:z-diffr-bpc}).
This means that the $W$ dependence of the diffractive cross section is steeper 
than for the total cross section, as expected in Regge theory.
In contrast, in the DIS domain at several \gevsq, the diffractive and total 
deep inelastic cross sections exhibit the same $W$ dependence, at variance 
with Regge theory expectations.
The value of $\alpha_\pom(0)$ for diffractive scattering is thus lower than 
for the total cross section (the latter is represented on the figure by the 
curve labelled ALLM, which corresponds to a Regge motivated parameterisation 
of the total $\gamma^\ast p$ cross section~\cite{ALLM}).

\subsubsection{Parton distributions.}

Parton distributions in the pomeron follow the DGLAP evolution equations, 
except for higher twist terms which can be significant, especially 
at large $\beta$ values, 
$\beta \gsim 0.7-0.8$~\cite{NNN,BEKW,DDISfactor}.

\begin{figure}[htbp]
\vspace{-0.cm}
\begin{center}
 \epsfig{file=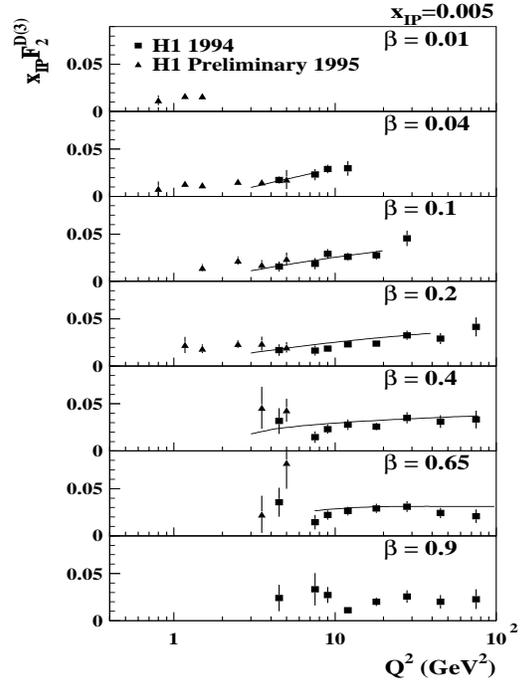,width=7cm,height=9cm}
\end{center}
\vspace{-0.5cm}
\caption{H1 measurement of the structure function $\xpom \cdot \fiidiii$
 for $\xpom = 0.005$ as a function of \qsq\ in bins of 
 $\beta$~\protect\cite{H1diffrvan}.
 The curves are the result of a DGLAP fit; they indicate the kinematical region
 over which the fit was performed.}
 \label{fig:scal-viol}
\end{figure}

Positive scaling violations are exhibited by DDIS at HERA, even for relatively
large values of $\beta$ (Fig.~\ref{fig:scal-viol}).
QCD fits performed by H1 provide parton distributions in the pomeron which 
are dominated by (hard) gluons at the starting scale $Q_{\rm o}^2 =3$ 
\gevsq~\cite{f2d3H1}~\footnote{
It should be stressed that only data up to $\beta = 0.65$ are used for
the DGLAP QCD fits. 
The details of the pomeron structure at higher $\beta$ values (e.g. the H1 
``peaked'' gluon or the H1 ``flat'' gluon distributions~\cite{f2d3H1}) are thus 
extrapolations outside the measurement domain and should not be taken too 
literally.}.

The ZEUS collaboration~\cite{Zjointfit} (and similarly the group~\cite{alvero})
has extracted the partonic content of the pomeron 
through a joint fit to the DDIS cross section, which probes the quarks 
directly, and diffractive jet photoproduction, which is mainly sensitive to the 
gluons. 
Although potentially sensitive to complications due to reinteractions
between the diffracted proton and remnants of resolved photons
(see below, section~\ref{sect:factbr}), these analyses confirm that most 
of the pomeron momentum is carried by gluons.

The pomeron pdf's extracted from QCD fits to inclusive DDIS can in turn
be convoluted with scattering amplitudes to describe specific processes.
This is performed using Monte Carlo simulations, in particular the Rapgap
model~\cite{rapgag}.
Several analyses of hadronic final states show a good agreement between
predictions and data~\cite{H1finalstate,ZEUSfinalstate}, which supports the 
universality of parton distributions in the pomeron.

\begin{figure}[htbp]
\vspace{-0.cm}
\begin{center}
 \epsfig{file=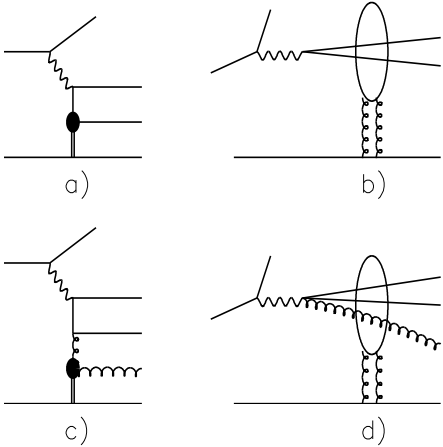,width=6cm,height=6cm}
\end{center}
\vspace{-0.5cm}
\caption{Deep inelastic diffractive scattering: left: the pomeron structure
function approach (Breit frame): a) quarkonic pomeron, no pomeron remnant; c)
gluonic pomeron, with a pomeron remnant; right: the photon Fock state approach
(proton rest frame): b) $q \bar q$ Fock state; d) $q \bar q g$ Fock state;
the pomeron is modelled as a two gluon system.}
\label{fig:approaches}
\end{figure}

The description of DDIS in terms of a partonic structure of the 
pomeron (Breit frame approach) can be complemented by an approach using the
proton rest frame (see Fig.~\ref{fig:approaches}).
In this approach, the photon is described as a superposition of
Fock states ($q \bar q$, $q \bar q g$, etc.), which are ``frozen'' 
during the hard interaction process~\cite{NNN,pech,BEKW,hebecker}.

At this conference, new results have been presented on two hard diffractive 
processes: dijet and charm production in DIS.
Hard diffraction has also been studied at HERA in the case of dijet
photoproduction~\cite{ZEUSjetsphotoprod,H1jetsphotoprod}.

\subsubsection{Diffractive dijet production.}

The H1 collaboration has measured diffractive dijet production with 
$p_T^{jet} > $ 4 GeV ($p_T$ is measured with respect to the $\gamma^{\ast} p$
axis), for DIS events with $4 < Q^2 < 80$ \gevsq\ and $\xpom < 0.05$.
A reasonable description of the differential distributions, both in 
normalisation and in shape, is obtained using pdf's extracted from inclusive 
DDIS~\cite{H1diffrdijets}.

\begin{figure}[htbp]
\vspace{-0.cm}
\begin{center}
 \epsfig{file=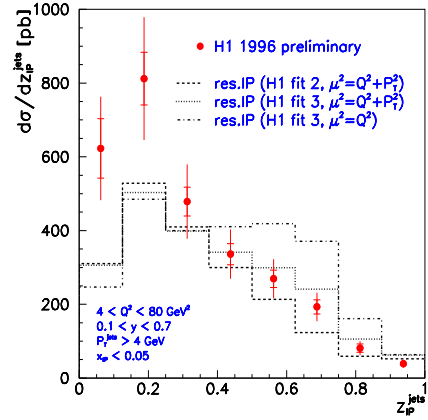,width=5.5cm,height=5.5cm}
\end{center}
\vspace{-0.3cm}
\caption{H1 measurement of the variable $\zpom$ for diffractive
 dijet production~\protect\cite{H1diffrdijets}. 
 The histograms represent predictions of the Rapgap 
 model~\protect\cite{rapgag} using pomeron pdf's extracted from inclusive DDIS: 
 the dashed and dotted histograms
 are for a ``flat'' gluon, with two different QCD scales; the dashed-dotted 
 histogram is for a ``peaked'' gluon~\protect\cite{f2d3H1}.}
 \label{fig:diffrdijets}
\end{figure}

Fig.~\ref{fig:diffrdijets} presents the distribution of the variable 
$\zpom = (M_{JJ}^2 + Q^2) / (M_X^2 + Q^2)$, where $M_{JJ}$ is the invariant 
mass of the two jet system; $\zpom$ represents the fraction of the pomeron 
momentum carried by the partons (gluons) entering the hard process.
In the absence of a pomeron remnant (Breit frame approach, 
Fig.~\ref{fig:approaches}a) or, equivalently, for a pure $q \bar q$ Fock state
of the photon (rest frame approach, Fig.~\ref{fig:approaches}b),
$M_{JJ} \simeq M_X$ and $\zpom \simeq 1$.
This is observed only for a small fraction of the data, as expected as a
consequence of ``colour transparency'': high $p_T$ jets correspond to a small
transverse distance between the quark and the antiquark, leading to mutual
screening into a colour neutral object which is thus not detected by the 
proton (Fig.~\ref{fig:approaches}b).
At variance, in the presence of an additional parton 
($q \bar q g$ or higher order Fock states, Fig.~\ref{fig:approaches}d), 
the parton pair leading to the jets is not in a colour singlet state and the
interaction with the proton takes place without attenuation due to colour
transparency.

\subsubsection{Diffractive charm production.}

Diffractive charm production in DIS has been studied both by the ZEUS and H1
collaborations in the channel $D^* \rightarrow K 2 \pi$, and by ZEUS
for  $D^* \rightarrow K 4 \pi$~\cite{ZEUSdiffrcharm,H1diffrcharm}.
The diffractive charm production rate is measured by ZEUS to be $\simeq$ 8\% 
of the total charm yield in DIS, and $\simeq$ 4\% for H1.
In view of the large errors, this corresponds only to a 2 $\sigma$ discrepancy.

\begin{figure}[htbp]
\vspace{-0.cm}
\begin{center}
 \epsfig{file=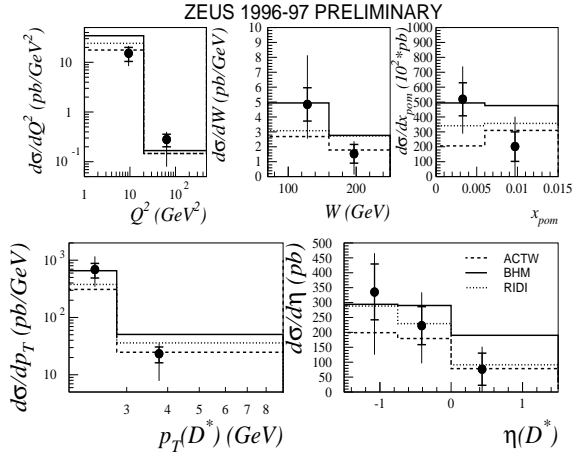,%
 bbllx=0.pt,bblly=20.pt,bburx=525.pt,bbury=407.pt,%
 width=7.5cm,height=6.cm}
\end{center}
\vspace{-0.3cm}
\vspace{0.cm}
\caption{ZEUS measurement of the diffractive $D^* \rightarrow K4\pi$
 cross section, as a function of \qsq, $W$, $\xpom$, the transverse momentum
 and the pseudorapidity of the $D^*$ particle~\protect\cite{ZEUSdiffrcharm}.
 The histograms represent predictions of different models.}
 \label{fig:diffrcharm}
\end{figure}

The shapes of the differential distributions are reproduced by calculations
including the pomeron pdf's extracted from inclusive DDIS (see
Fig.~\ref{fig:diffrcharm}).
As in the case of jet diffractive production, the absence of a peak close 
to 1 in the $\zpom$ distribution (not shown, H1 analysis~\cite{H1diffrcharm}) 
is attributed to a dominant role of $q \bar q g$ or higher order Fock states, 
due to the effect of colour transparency.

\subsubsection{Conclusions.}

The HERA experiments have provided a rich sample of results on diffractive 
processes in the presence of a hard scale (diffractive final state studies 
in DIS, jet and charm production).
Within the limits of the present statistics, these data are consistent with 
the universality of the pdf's extracted from QCD fits to inclusive DDIS.

\subsection{Hard diffraction at the Tevatron.}

Even before HERA data taking, hard diffractive processes were observed at
the CERN $p \bar p$ collider by the UA8 experiment~\cite{UA8}:
while the diffractively scattered proton was detected in a proton spectrometer,
high $p_T$ jets were reconstructed in the central detector.
This observation supported the hypothesis of a partonic component of
diffraction~\cite{Ing-schlein}.

At the Tevatron collider, hard diffraction is being extensively studied by the
D0 and CDF collaborations~\cite{snow}, which complements the studies at HERA.

\subsubsection{Single diffraction, double diffraction and double 
                  pomeron exchange.}

\begin{figure}[htbp]
\vspace{-0.cm}
\begin{center}
 \epsfig{file=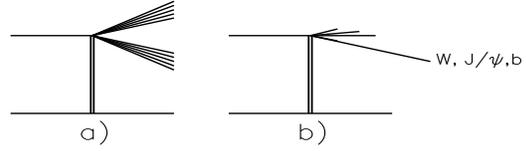,width=7cm,height=2cm}
\end{center}
\vspace{-0.3cm}
\caption{Hard single diffraction at the Tevatron: a) dijet 
 production (gap/$\bar p$ + 2 jets); b) $W$, $J/\psi$, $b \bar b$ 
 production.}
 \label{fig:tevsingle}
\end{figure}

Hard single diffraction processes are studied at the Tevatron through the
production of high $p_T$ jets~\cite{CDFSDjets,D0jets} 
(Fig.~\ref{fig:tevsingle}a), and of 
$W$ bosons~\cite{CDFW}, $J/\psi$ mesons~\cite{borras} and  
$b$ particles~\cite{CDFb} (Fig.~\ref{fig:tevsingle}b).
These events are identified either through the detection of the diffractively
scattered $\bar p$ in a proton spectrometer (CDF dijet 
events), or by the presence of a gap in pseudorapidity, devoid of hadronic 
activity, in the calorimeter and the tracking detector.
Production rates are at the 1\% level compared to the corresponding
non-diffractive processes~\cite{Tevrates}.

\begin{figure}[htbp]
\vspace{-0.cm}
\begin{center}
 \epsfig{file=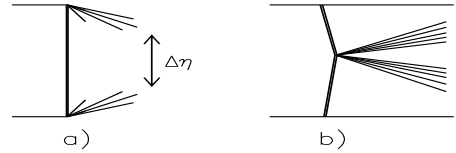,width=6cm,height=2cm}
\end{center}
\vspace{-0.3cm}
\caption{a) Dijet production with a gap in rapidity (jet + gap + jet),
 attributed to colour singlet exchange;
 b) dijet production by double pomeron exchange (gap/$\bar p$ + 2 jets + gap).}
 \label{fig:tevdouble}
\end{figure}

Hard double diffraction (see Fig.~\ref{fig:tevdouble}a) is studied through 
the production of two jets separated by a gap in rapidity attributed to colour 
singlet exchange~\cite{DDjets,D0jets}.
The rate of such events has been studied for $\sqrt{s} = 630$ and for  
$\sqrt{s} = 1800$.
The ratio $R_{630/1800}$ is measured to be $2.4 \pm 0.9$ by CDF and $1.9 \pm
0.2$ by D0.
This decrease of the diffractive process with increasing energy is at 
variance with expectations based on simple BFKL evolution~\cite{delduca-tang}, but 
is predicted by models of soft colour recombination~\cite{eboli}.

Finally hard dijet production has also been observed in the central detectors 
for events containing a scattered $\bar p$ identified in the proton 
spectrometer and a rapidity gap on the other side of the detector 
(CDF)~\cite{borras}, or a rapidity gap on each side of the detector 
(D0)~\cite{D0jets} (Fig.~\ref{fig:tevdouble}b).
These events are found to be produced at the $10^{-4}$ level of the
corresponding non-diffractive process, which is consistent with a picture of
double pomeron exchange, each pomeron exchange corresponding to a probability 
at the 1\% level.

\subsubsection{Factorisation breaking.}
                                            \label{sect:factbr}

Following a procedure similar to ZEUS~\cite{Zjointfit}, the CDF collaboration 
has determined
the partonic content of the pomeron by taking advantage of the different 
sensitivities of the various processes (dijet, $W$ and $b$ production) to  
quarks and gluons~\cite{CDFb}.
The production rates were compared to predictions of 
the model Pompyt~\cite{pompyt}, which is based on the assumption of a 
factorisable pomeron flux; a hard partonic content of the pomeron was assumed. 

\begin{figure}[htbp]
\vspace{-0.cm}
\begin{center}
 \epsfig{file=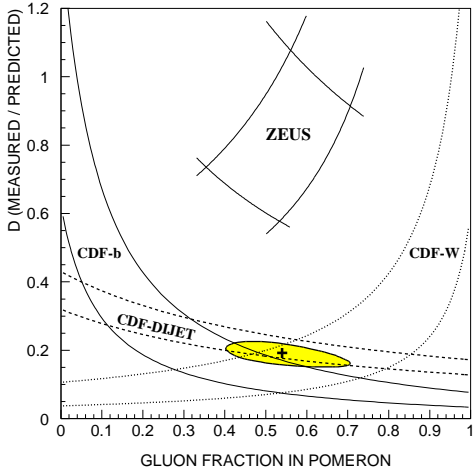,%
 bbllx=5pt,bblly=10pt,bburx=458pt,bbury=462pt,%
 width=6.2cm,height=6.2cm}
\end{center}
\vspace{-0.3cm}
\caption{Ratio of measured to predicted diffractive rates as a function of 
 the gluon content of the pomeron, for CDF dijet, $W$ and $b$ production
 and for a measurement by ZEUS of DDIS and diffractive jet 
 photoproduction. 
 The predictions are from the Pompyt model~\protect\cite{pompyt} with a hard 
 pomeron structure.
 The shaded area is the 1 $\sigma$ contour of a fit to the three CDF 
 results~\protect\cite{CDFb}.}
 \label{fig:factbreak}
\end{figure}

A gluon fraction of $0.55 \pm 0.15$ is found, which is consistent with 
measurements at HERA (see Fig.~\ref{fig:factbreak}), but the measured rates at 
the Tevatron are significantly lower than expected, the reduction factor being 
$D = 0.18 \pm 0.04$, whereas the order of magnitude of the HERA results is 
reproduced. 

Similarly, predictions for the diffractive production rate of dijets and $W$
bosons~\cite{alvero} and for charm production and double pomeron 
exchange~\cite{alvero2} based on pomeron pdf's extracted from inclusive DDIS
indicate that factorisation, which is verified in DIS, is broken in the
case of diffractive hadron--hadron interactions.

The factorisation breaking is quantified in terms of a ``survival probability''.
In hadron--hadron scattering, additional interactions between the diffractively
scattered particle and remnants of the other beam particle can destroy 
the rapidity gap, whereas this effect is absent in DDIS~\cite{survival_proba}.
This leads to a reduction of the diffractive rates at the Tevatron compared 
to predictions based on HERA DDIS data~\footnote {
In the case of diffractive photoproduction at HERA, additional interactions 
can also take place between the scattered proton and the resolved components 
of the photon. 
An indication for such an effect has been found in diffractive dijet
photoproduction by H1~\cite{H1jetsphotoprod}.}.
The energy dependence of the gap survival probability could also explain the
observed decreasing rate of colour singlet exchange between jets for 
increasing $\sqrt {s}$.

\subsubsection{Conclusion.}

In conclusion, active studies of hard diffraction are performed at the Tevatron,
where diffractive processes represent about 1\% of the corresponding
non-diffractive processes.
However, naive calculations for diffraction rates at the Tevatron based on 
pomeron pdf's obtained at HERA do not describe the data, which are about a 
factor 4 lower.
This reduction of the gap survival probability could be attributed to underlying
interactions between beam particle remnants.

\subsection{Exclusive production of vector particles at HERA.}

Numerous vigorous attempts are being made to use pQCD to calculate the cross 
section for several diffractive processes at HERA.
Among them, diffractive (exclusive) production of a vector particle, either 
a photon or a vector meson, provides the most solid theoretical ground, as well
as numerous high quality data.
We concentrate here on the new results presented at this conference.

\subsubsection{Deeply virtual Compton scattering.}

\begin{figure}[htbp]
\vspace{-0.cm}
\begin{center}
 \epsfig{file=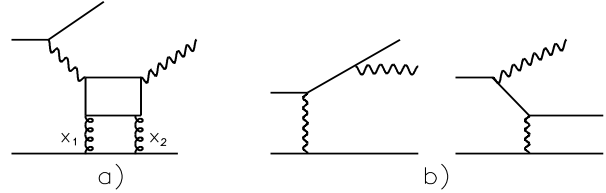,width=8.cm,height=2.6cm}
\end{center}
\vspace{-0.3cm}
\caption{a) The DVCS process; b) the two LO diagrams contributing to the 
 Bethe-Heitler (QED Compton) process.}
 \label{fig:dvcsdiag}
\end{figure}

Deeply virtual Compton scattering (DVCS): $e + p \rightarrow e + p + \gamma$
(see Fig.~\ref{fig:dvcsdiag}a) is a gold-plated process for the study of pQCD in 
diffraction~\cite{dvcsth}. 
At high \qsq, the process is completely perturbatively calculable, since the
incoming and outgoing photon wave functions and all couplings are known, 
and no strong interactions between final state particles affect the
calculation.

To extract the DVCS cross section, account has to be taken of the interference
with the Bethe-Heitler (QED Compton) process (Fig.~\ref{fig:dvcsdiag}b), 
but the two processes correspond to different regions of phase-space.
The DVCS process is dominated by cases where the photon is emitted in the proton
direction, since the photon flux factor in the electron is $\propto 1/y$, 
whereas for the Bethe-Heitler process, the photon is
dominantly emitted in the electron direction.

The ZEUS collaboration~\cite{dvcsZEUS} has for the first time at this 
conference shown evidence for the DVCS process, obtained with a
sample of DIS events with $Q^2 > 6$ \gevsq\ containing an 
electromagnetic cluster with energy larger than 10 GeV emitted in the backward
region of the detector, a second electromagnetic cluster with energy 
larger than 2 GeV detected in the central region, at most one reconstructed 
track, and a maximum of 0.5 GeV additional energy reconstructed in the detector.

%
\begin{figure}[htbp]
\begin{center}
\setlength{\unitlength}{1.0cm}
\begin{picture}(8.0,5.3)
\put(1.5,0.0){\epsfig{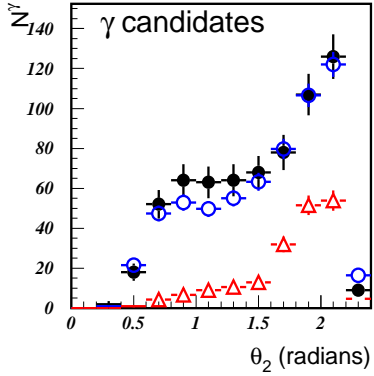}}
\end{picture}
\vspace{-0.3cm}
\caption{ZEUS measurement of the polar angle distribution of the photon 
 candidate with energy larger than 2 GeV for $e p \gamma$ 
 events~\protect\cite{dvcsZEUS}.  
 The data are the full dots, the predictions for the Bethe-Heitler process
 are the open triangles, and the predictions of a DVCS + Bethe-Heitler
 simulation are the open circles.}
 \label{fig:dvcsexp}
\end{center}
\end{figure}

Fig.~\ref{fig:dvcsexp} shows the polar angle distribution of the second 
cluster, when identified as a photon
The excess of events over the Bethe-Heitler prediction is consistent, 
in shape and normalisation, 
with the predictions of a simulation aimed at describing the DVCS and 
Bethe-Heitler processes, including the interference term.

It should be noted that, in the DVCS process, an incoming virtual photon is
converted into a real photon. 
Kinematics imply that longitudinal momentum must be transferred to the
proton, and the two gluons are thus not emitted and reabsorbed with the same
energy ($x_1 \neq x_2$).
This observation has led to the concept of ``skewed parton
distributions''~\cite{SPD}.
The DVCS process is an ideal tool to study correlations between gluons in
the proton.


\subsubsection{Vector Meson Production.}

Vector meson (VM) production, both in photo- ($Q^2 \simeq 0$) and
electroproduction has been intensively studied at HERA, for $\rho, \omega,
\phi, \rho^{\prime}, J/\psi, \Psi^{\prime}, 
\Upsilon$~\cite{marage,cibo,clerbaux,rhoh1,rhozeus,phizeus,jpsih1}.

\begin{figure}[htbp]
\vspace{-0.cm}
\begin{center}
 \epsfig{file=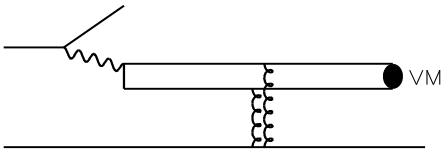,width=6.cm,height=2.cm}
\end{center}
\vspace{-0.3cm}
\caption{Vector meson production at high energy.}
 \label{fig:vmdiag}
\end{figure}

These processes can be computed as the
convolution of three amplitudes involving very different time scales in the
proton rest frame (see Fig.~\ref{fig:vmdiag}): the 
$\gamma \rightarrow q \bar q$ transition (a long distance process at high 
energy), the hard scattering of the $q \bar q$ pair (a short time process) 
and the $q \bar q \rightarrow VM$ recombination (on a typical hadronic scale 
of 1 GeV, boosted to the proton rest frame).

\begin{figure}[htbp]
\vspace{-0.cm}
\begin{center}
 \epsfig{file=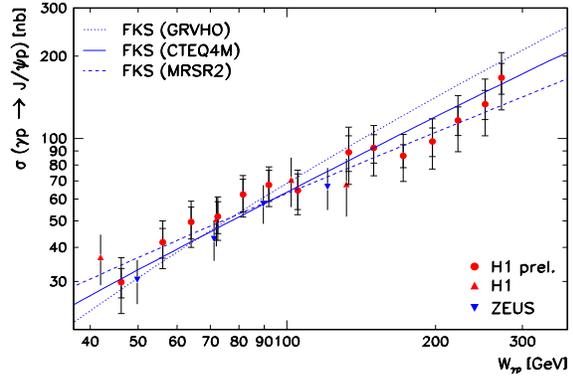,%
 bbllx=22.pt,bblly=28.pt,bburx=516.pt,bbury=377.pt,%
 width=7.5cm,height=5.cm}
\end{center}
\vspace{-0.3cm}
\caption{Energy dependence of the $J/\psi$ photoproduction cross section at 
 HERA~\protect\cite{jpsih1}, compared to QCD predictions~\protect\cite{fks} 
 using several pdf's (the absolute normalisations have been adjusted to 
 the data).}
 \label{fig:jpsi}
\end{figure}

\vspace{0.3cm}
{\bf Energy dependence of the cross section.}
In the presence of a hard scale (large photon virtuality, heavy quark mass,
large $|t|$), the hard process amplitude is modelled as two gluon exchange
(reggeised gluons in a BFKL approach).
The cross section is expected to be proportional to $|x G(x)|^2$ and thus 
exhibit a ``hard'' energy dependence, which is clearly observed in the case 
of $J/\psi$ 
photoproduction (Fig.~\ref{fig:jpsi}): the cross section can be 
parameterised, in a Regge inspired form, as
$\sigma(\gamma^\ast p) \propto W^{4 \alpha_{\pom} (\bar t) - 4}$,
with $\alpha_{\pom}(\bar t ) \simeq 1.20$, and QCD predictions~\cite{fks}
describe the data well.
For light VM production ($\rho, \phi$),  $\alpha_{\pom}(0)$ is
observed to increase from a ``soft'' value typical of hadron--hadron 
scattering in photoproduction, to a value suggestive of a ``hard'' behaviour 
at high \qsq\ (see e.g.~\cite{rhoh1}).

\vspace{0.3cm}
{\bf \boldmath{$Q^2$} dependence of the cross section.}
The \qsq\ dependence of the cross section for electroproduction of $\rho$ 
mesons can be parameterised in the form 
$\sigma(\gamma^\ast p) \propto 1 / (Q^2 + M^2_\rho)^n$, 
with $n = 2.3 \pm 0.1$~\cite{rhoh1}.
This behaviour is consistent with pQCD calculations 
($\propto 1 / Q^6$)~\cite{brodsky}, when account is taken of the \qsq\
dependence of $xG(x)$ and \as.

\begin{figure}[htbp]
\setlength{\unitlength}{1.0cm}
\begin{center}
\begin{picture}(8.0,8.0)
\put(0.0,0.){\epsfig{file=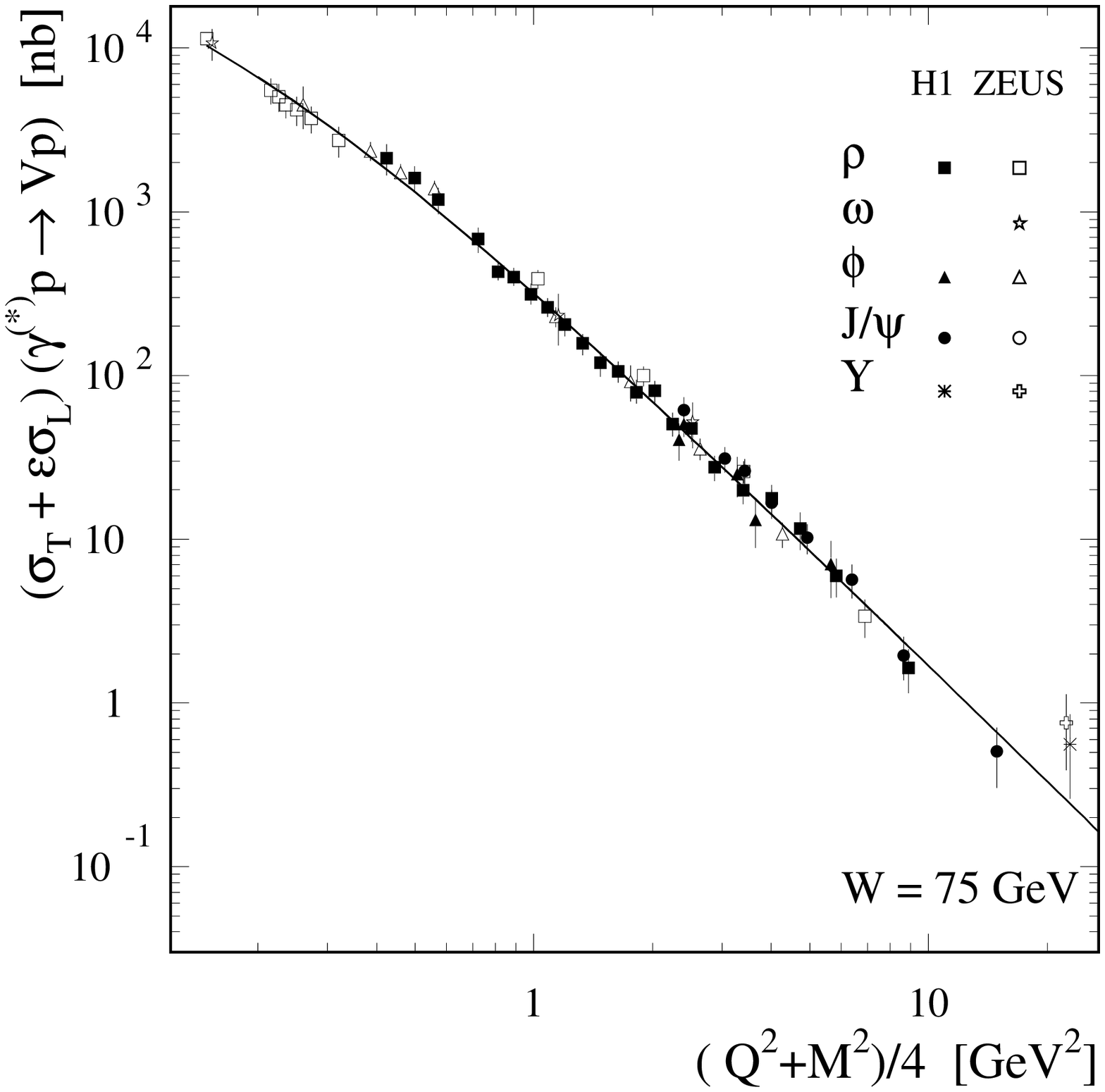,width=8cm,height=8cm}}
\put(1.4,1.3){\epsfig{file=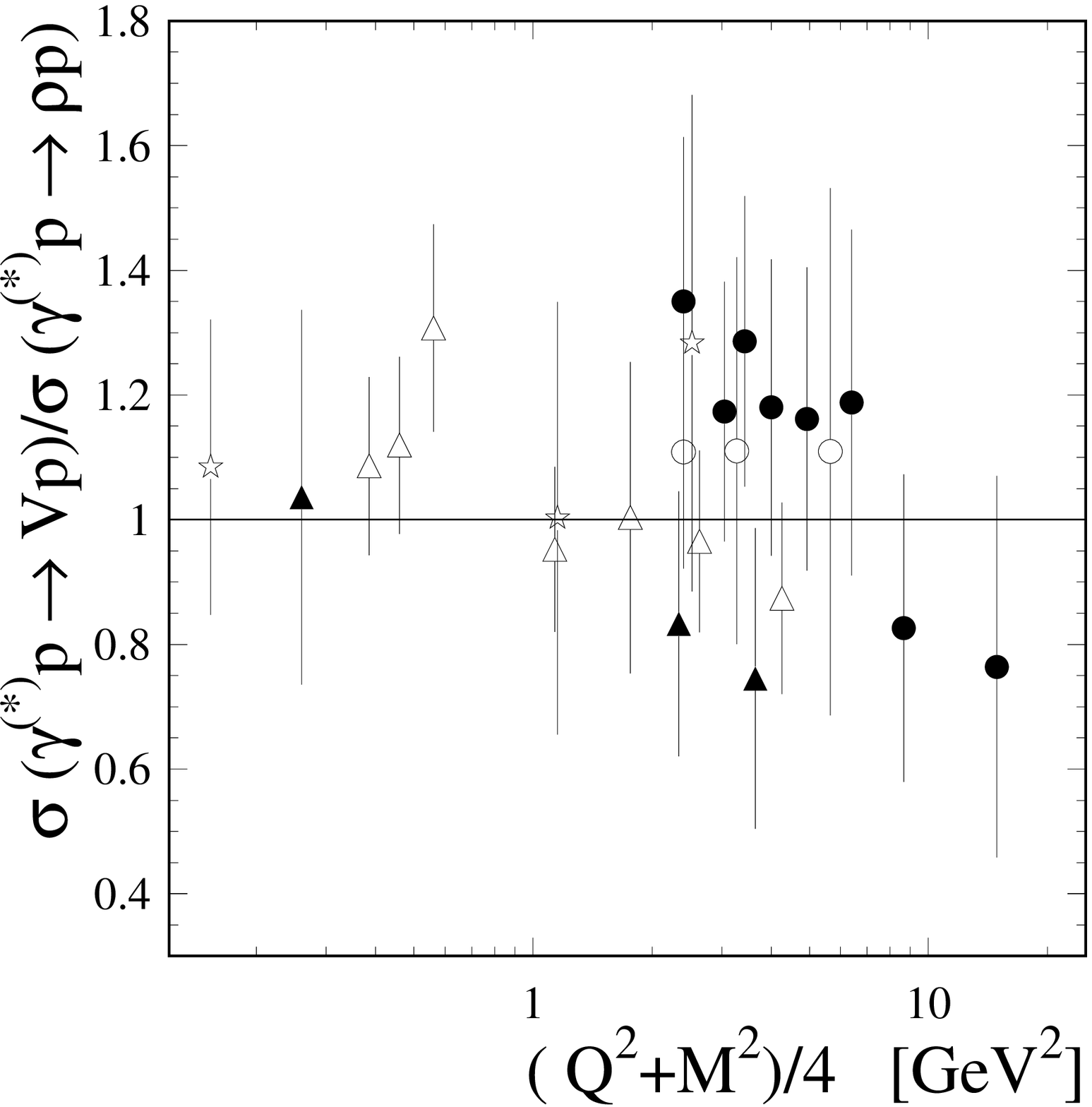,width=3.75cm,height=3cm}}
\end{picture}
\caption{Cross section for elastic vector meson electroproduction, as a function
 of the variable $\frac {1}{4} (Q^2 + M_V^2)$~\protect\cite{clerbaux}.}
 \label{fig:baq2plusm2}
\end{center}
\end{figure}

The ratio of cross sections for $\phi$ and \jpsi\ to $\rho$ meson 
electroproduction~\cite{rhozeus} increases significantly with \qsq, towards 
values compatible
with the quark counting rule (respectively the ratios 2/9 and 8/9), 
convoluted with the effects of wave functions~\cite{fks}.
It is interesting to note that, when plotted as a function of the variable 
$\frac {1}{4} (Q^2 + M_V^2)$, all $\gamma^\ast p \rightarrow VM \ p$ 
cross sections exhibit
a common behaviour (see Fig.~\ref{fig:baq2plusm2}~\cite{clerbaux})~\footnote{
In the case of $\Upsilon$ production, large effects of skewed parton
distributions may have to be taken into account~\cite{upsilonspd}.}.

\vspace{0.3cm}
{\bf \boldmath{$t$} dependence of the cross section.}
The $t$ dependence of the cross section for vector meson elastic production
can be parameterised for low $|t|$ ($|t| \lsim 1-2$ \gevsq) as 
${\rm d} \sigma / {\rm d} t \propto e^{-b |t|}$,
the slope parameter $b$ being related to the transverse size of the interacting
objects: $b \simeq R_p^2 + R_{VM}^2 + R_{\pom}^2$.
In Regge theory, the $t$ distribution is expected to shrink with energy as
$b(s) = b(s_{{\rm o}}) + 2 \alpha^\prime \cdot \ln (s / s_{{\rm o}})$,
with the trajectory slope $\alpha^\prime_\pom \simeq 0.25$ \gevsqm.
At high energy, little shrinkage is expected in QCD (BFKL evolution),
since $\alpha^\prime_{BFKL}$ is expected to be small~\cite{brodsky-BFKL-NLO}.

\begin{figure}[htbp]
\vspace{-0.cm}
\begin{center}
 \epsfig{file=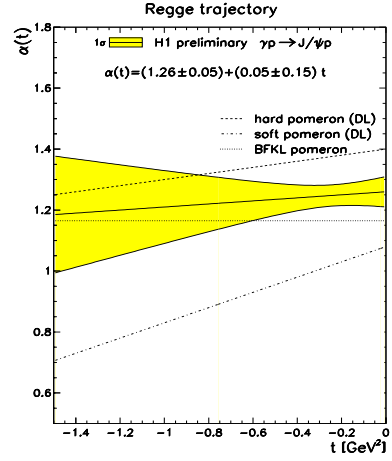,width=5.cm,height=6.cm}
\end{center}
\vspace{-0.3cm}
\caption{The Regge trajectory 
 $\alpha(t) = \alpha(0) + \alpha^\prime \cdot t$ measured by H1 for \jpsi\ 
 photoproduction (full line, the error being given by the
 shaded area)~\protect\cite{jpsih1}.
 The BFKL prediction~\protect\cite{brodsky-BFKL-NLO} is shown as the dotted 
 line, and the ``soft'' pomeron is the dashed-dotted line; the dashed line 
 corresponds to a hard pomeron model~\protect\cite{DLtwopom}.}
 \label{fig:shrinkage}
\end{figure}

A measurement of the evolution of the $t$ distribution as a function of $W$ 
within one experiment, H1, has been presented for the first time at this 
conference for \jpsi\ photoproduction~\cite{jpsih1}. 
In spite of large errors, the slope of the trajectory 
$\alpha^\prime = 0.05 \pm 0.15$ \gevsqm\ is found to be 
consistent with 0 (Fig.~\ref{fig:shrinkage}), which supports the QCD expectation. 

\vspace{0.3cm}
{\bf Polarisation.}
Detailed studies have been performed of the polarisations state of 
$\rho$~\cite{rhoh1,rhozeus} and $\phi$~\cite{phizeus} mesons, particularly 
in electroproduction.
Although $s$-channel helicity conservation (SCHC) is dominantly observed to 
hold, a small but significant spin flip amplitude is measured in the 
transition from a transverse photon to a longitudinal $\rho$ meson, at the 
level of $8 \pm 3$\%; the longitudinal to transverse transition and the 
double flip amplitude are compatible with 0 within errors~\cite{clerbaux}.
These features are qualitatively reproduced by QCD based 
calculations~\cite{vmmodels}.

\begin{figure}[htbp]
\vspace{-0.cm}
\begin{center}
 \epsfig{file=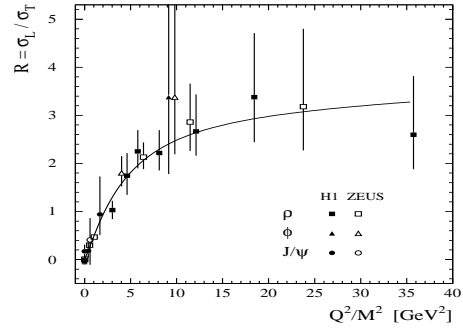,width=6.cm,height=4.3cm}
\end{center}
\vspace{-0.3cm}
\caption{Ratio $R = \sigma_L / \sigma_T$ of the longitudinal to transverse 
 cross sections for $\rho$, $\phi$ and \jpsi\ meson production,
 as a function of $Q^2 / M^2$~\protect\cite{clerbaux}.
 The curve represents a phenomenological fit.}
 \label{fig:R}
\end{figure}

The ratio $R = \sigma_L / \sigma_T$ of the longitudinal to the transverse cross
section has been measured for $\rho$, $\phi$ and \jpsi\ meson production, and
found to increase with \qsq\ in the DIS region (Fig.~\ref{fig:R}).
Although this increase is slower than anticipated~\cite{brodsky},
it is reproduced by some models based on QCD~\cite{vmmodels} or on generalised 
vector meson dominance (GVDM)~\cite{sss}.
When plotted as a function of the quantity $Q^2 / M^2$~\cite{clerbaux}, the 
measurements for the different vector mesons appear to follow a common behaviour 
(Fig.~\ref{fig:R}).


\section{Indications for non-linear effects ?}

The numerous results presented in this review provide a bright support for the
presently available QCD calculations: impressive tests of the DGLAP evolution 
in DIS over a huge kinematic domain, 
indications for the relevance of the BFKL evolution in several channels at 
very high energy, 
relevance of the QCD approach for understanding diffraction and for exclusive
vector particle production.

However, several intriguing features, both in inclusive DIS and in diffraction, 
suggest that this picture might have to be complexified.
They are discussed in ref.~\cite{mueller,glm,koppovh,kgb}, where it is advocated
that they could be related to a very large density of 
partons at very low $x$ and at \qsq\ of the order of a few \gevsq, leading to
saturation effects and a breakdown of the DGLAP and BFKL linear evolution 
equations.
Unitarity constraints~\cite{amaldi,koppovh} play an essential role in this 
dynamics.

\begin{figure}[htbp]
\vspace{-0.cm}
\begin{center}
 \epsfig{file=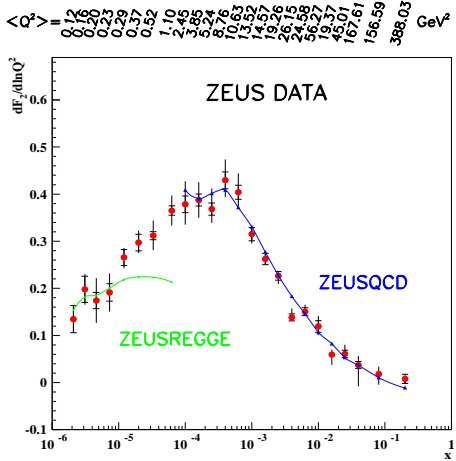,%
 bbllx=45.pt,bblly=170.pt,bburx=540.pt,bbury=687.pt,%
 width=6.cm,height=6.cm}
\end{center}
\vspace{-0.3cm}
\caption{Logarithmic derivative of the $F_2$ structure function measured by
 ZEUS, as a function of $x$~\protect\cite{Z-pdf}; the corresponding 
 average value of \qsq\ is also indicated.
 The curves correspond to a NLO DGLAP fit and to a Regge parameterisation.}
 \label{fig:caldwell}
\end{figure}

In DIS, it is observed that the parton 
distributions extracted from (statistically satisfactory) DGLAP fits
to the measured total cross section
exhibit an unexpected behaviour at low \qsq: the gluon density at very low 
$x$ becomes very small, even possibly negative, and the sea quark density
is larger than for the gluon, whereas at larger \qsq\ the gluon density drives
the sea behaviour (see Fig.~\ref{fig:Z_gluon}).
In addition, the logarithmic derivative ${\rm d} F_2 / {\rm d} \ln Q^2$ of the 
$F_2$ structure function, presented in Fig.~\ref{fig:caldwell} as a function 
of $x$ and the corresponding average value of \qsq, shows an unexpected 
turn over at low $x$ and $Q^2 \simeq$ a few \gevsq.
Such a turn over is not observed at higher $x$ for the same \qsq\ range, 
suggesting that it is not due to higher twist effects.

\begin{figure}[htbp]
\vspace{-0.cm}
\begin{center}
 \epsfig{file=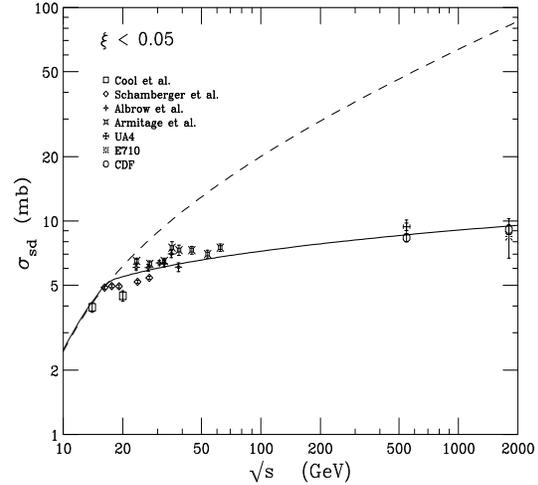,%
 bbllx=34.pt,bblly=289.pt,bburx=604.pt,bbury=652.pt,%
 width=7.cm,height=5.cm}
\end{center}
\vspace{1.0cm}
\caption{Total single diffraction cross section for 
  $ p (\bar p) + p \rightarrow p (\bar p) + X$ as a function of 
  $\protect\sqrt{s}$, compared to predictions from a Regge extrapolation of 
  the low energy data (dashed line).
  The solid line describes a phenomenological model~\protect\cite{dino}.}
 \label{fig:dino}
\end{figure}

In diffractive DIS, the total cross section is observed to present a ``hard''
behaviour (see section~\ref{sect:endep-f2d3} and Fig.~\ref{fig:z-diffr-bpc}),
whereas the expectations are that the dominant topology would correspond to 
the ``aligned jet model'', with small $p_T$ values and a ``soft'' energy 
dependence similar to that of hadron--hadron scattering.
In soft hadronic diffractive dissociation  
$ p (\bar p) + p \rightarrow  p (\bar p) + X$,
the measured cross section at high energy (CERN and Tevatron colliders) is 
significantly lower than expected from Regge theory (Fig.~\ref{fig:dino}).
Finally, as discussed in section~\ref{sect:factbr}, hard diffractive events 
at the Tevatron are suppressed compared to expectations based on inclusive DIS
measurements.
All these features are also attributed to very high parton densities and 
saturation effects.

\section{Conclusions.}

In conclusion, huge amounts of data have been presented at this conference
about hadron structure, low $x$ physics and diffraction.
The progress in these domains is impressive, both on the theoretical and the
experimental sides.
The parton distributions in the proton are precisely measured over 
most of the $x$ domain, and new measurements are being performed.
The $ep$ total cross sections are described with high precision by the DGLAP 
evolution equations over a huge kinematic domain, but indications for the
relevance of the BFKL evolution begin to appear in exclusive channels.
A description of the pomeron in terms of partonic structure functions gives a
consistent picture of the data in DIS at HERA, which is complemented by 
perturbative QCD calculations for hard processes.
Hard diffraction is also intensively studied at the Tevatron in several
channels.
Finally, at HERA, the DVCS process and vector meson production, with a large 
amount of detailed data, provide a clean laboratory for a perturbative 
QCD understanding of diffraction.
Intriguing features however suggest that the linear DGLAP and BFKL evolution
equations might not be sufficient to describe all data, with possibly an 
indication for saturation effects at very low $x$ and low \qsq\ values.


\section*{Acknowledgements}
I wish to thank numerous colleagues for their help in preparing this 
talk, 
in particular H.~Abramowicz, K.~Borras, B.~Clerbaux, E.A.~De~Wolf, E.~Elsen, 
M.~Erdmann, A.~Garfagnini, A.~Goussiou, A.~Martin, P.~Merkel, C.~Royon, 
S.~Schlenstedt, U.~Stoesslein, G.~Snow. 
I also warmly thank B.~Clerbaux for her useful comments to the text of
these proceedings.



\end{document}